\documentclass[12pt]{article}
\pdfoutput=1
\usepackage{multirow}
\usepackage{calc}
\usepackage{amsmath,amssymb,amsthm,amscd}
\numberwithin{equation}{section}
\usepackage[bf]{caption}
\usepackage{longtable}
\usepackage{array}
\usepackage{enumerate}
\usepackage{float}
\usepackage{subfig}
\usepackage{multicol}
\usepackage{multirow}
\usepackage{epsfig}
\usepackage{graphicx}
\usepackage{pict2e}
\usepackage{color}
\usepackage{authblk}
\usepackage{cite}
\usepackage[all]{xy}
\usepackage[width=1.09\textwidth]{caption}
\usepackage{bbm}
\usepackage{hyperref}

\usepackage[utf8]{inputenc}
\usepackage[T1]{fontenc}
\usepackage[english]{babel}
\usepackage{xspace}
\usepackage{pifont}
\usepackage[OT2,T1]{fontenc}

\DeclareSymbolFont{cyrletters}{OT2}{wncyr}{m}{n}
\DeclareMathSymbol{\Sha}{\mathalpha}{cyrletters}{"58}

\makeatletter
\newcommand\xleftrightarrow[2][]{%
  \ext@arrow 9999{\longleftrightarrowfill@}{#1}{#2}}
\newcommand\longleftrightarrowfill@{%
  \arrowfill@\leftarrow\relbar\rightarrow}
\makeatother

\newcommand{\beq}{\begin{equation}}
\newcommand{\eeq}{\end{equation}}
\newcommand{\bea}{\begin{eqnarray}}
\newcommand{\eea}{\end{eqnarray}}

\begin{document}

\baselineskip=15pt
\begin{titlepage}

\begin{center}
\vspace*{ 2.0cm}
{\Large {\bf F-theory on Quotients of Elliptic Calabi-Yau Threefolds}}\\[12pt]
\vspace{-0.1cm}
\bigskip
\bigskip 
{
{{Lara B.~Anderson}$^{\,\text{a,b}}$},  {{James~Gray}$^{\,\text{a,b}}$} and {{Paul-Konstantin~Oehlmann}$^{\,\text{a,b}}$}
\bigskip }\\[3pt]
\vspace{0.cm}
{\it 
 ${}^{\text{a}}$ Physics Department,~Robeson Hall,~Virginia Tech,~Blacksburg,~VA 24061,~USA   \\
 ${}^{\text{b}}$ Simons Center for Geometry and Physics,~Stony Brook,~NY 11794,~USA
}
\\[2.0cm]
\end{center}

\begin{abstract}
\noindent
In this work we consider quotients of elliptically fibered Calabi-Yau threefolds by freely acting discrete groups and the associated physics of F-theory compactifications on such backgrounds. The process of quotienting a Calabi-Yau geometry produces not only new genus one fibered manifolds, but also new effective $6$-dimensional  physics. These theories can be uniquely characterized by the much simpler covering space geometry and the symmetry action on it. We use this method to construct examples of F-theory models with an array of discrete gauge groups and non-trivial monodromies, including an example with $\mathbb{Z}_6$ discrete symmetry. 
\end{abstract}

\end{titlepage}
\clearpage
\setcounter{footnote}{0}
\setcounter{tocdepth}{2}
\tableofcontents
\clearpage
 
 \section{Introduction}
 \label{sec:intro}
 Compactifications of F-theory provide a powerful tool in the study and classification of strongly coupled 6-dimensional theories, including 6- (and 5-dimensional) superconformal field theories (SCFTs) \cite{Heckman:2013pva,DelZotto:2014fia,DelZotto:2014hpa,Heckman:2015bfa,Jefferson:2017ahm}. To this end, the structure of elliptically fibered Calabi-Yau (CY) geometries and their singular degenerations is of clear relevance. In recent work \cite{Anderson:2018heq}, compactifications of F-theory on non-simply connected CY manifolds and their physical implications were studied. Such CY fibrations exhibit multiple fibers (i.e. everywhere singular fibers) and notably, the presence of discretely charged ``superconformal matter.'' More precisely, discretely charged matter is found to be localized on singular loci within the base manifold of the fibration where orbifold-type singularities are located. Over such points in the base, the CY fibration develops multiple fibers\footnote{Notably, multiple fibers have played an important role before in the context of quotient theories in the context of CHL \cite{deBoer:2001wca} and little string theories \cite{Bhardwaj:2015oru}.}. Within the resulting effective field theory, these singularities correspond to ``strongly coupled sectors'' which become SCFTs in the limit that gravity is decoupled.
 
 In this paper, we take a more systematic look at such CY quotients, extending previous work \cite{Anderson:2018heq}. In particular, we demonstrate that it is possible to systematically characterize the effective field theories resulting from F-theory compactifications on CY quotients that are obtained by freely acting discrete symmetries acting on covering spaces that are \emph{elliptically fibered -- i.e. torus fibered with section}.
 
Consider such a quotient of a smooth Calabi-Yau threefold, $X$, by a freely acting discrete symmetry, $\Gamma$. It has been demonstrated (see e.g. \cite{Donagi:1999ez}) that if the symmetry $\Gamma$ leads to a resulting CY threefold, $\tilde{X}=X/\Gamma$ which is also torus fibered ($\tilde{\pi}: \tilde{X} \rightarrow \tilde{B}$), then this new geometry will be a \emph{genus one fibered} manifold (i.e. admitting only multi-sections but no true sections to the fibration). We will frequently refer to the covering space ($X)$ as the ``upstairs'' geometry (or in an abuse of notation, sometimes refer to the associated physics as the ``upstairs theory'') and to quotient manifold $\tilde{X}$ as the ``downstairs'' geometry (or theory). In this work we will present a direct way of calculating the F-theory effective physics associated to such a geometry, including the degrees of freedom associated to singularities in the base manifold, $\tilde{B}$ (which lead to an ${\cal A}_n$ $(2,0)$ superconformal theory in the decoupling limit, with discretely charged superconformal matter) using only the geometry of the covering space $X$ and the symmetry action thereon.

In the case of genus one fibered manifolds, it is usually a somewhat difficult process to extract the F-theory effective physics from a compactification of a fibered manifold without a section\cite{Braun:2014oya,Anderson:2014yva,Mayrhofer:2014haa,Mayrhofer:2014laa,Morrison:2014era,Cvetic:2015moa}. In particular, the process of writing down the physical theory is usually accomplished by describing the Jacobian of the fibration \cite{artin_tate} (which does admit a holomorphic section). However, the practical construction of Jacobians of CY threefolds is not known in general. Moreover, the strength of this construction is frequently justified by considering the \emph{dynamical} connection of a genus-one fibered manifold and its Jacobian within an M-theory limit. However, only in some cases is it known how to explicitly and dynamically connect the elliptically fibered Jacobian manifold with the original multisection fibration via conifold-type transitions \cite{Morrison:2014era,Anderson:2014yva}.

By contrast, here we utilize the theorem of Shioda, Tate and Wazir \cite{shioda,shioda2,COM:213767} to categorize the divisors of $X$ into those that are horizontal (i.e. sections) versus vertical (i.e. pull-backs from divisors in the base), or fibral divisors associated to non-Abelian gauge symmetries. By studying the action of the symmetry $\Gamma$ on a set of effective divisors in this set, we will derive a series of simple rules which will allow us to characterize the effective theory defined on the quotient manifold and verify that the associated massless spectrum in the $6$-dimensional theory is consistent with anomalies.

It should be stressed here that in the examples presented in in this work, frequently the geometry of the genus-one fibered CY manifold and its Jacobian can differ substantially\footnote{We thank M. Esole, A. Grassi, and S. Katz for helpful conversations on this point.} (for instance $h^{1,1}(\tilde{X})< h^{1,1}(Jac(\tilde{X})_{resolved})$) and as a result, care must be taken in the context of considering the effective physics of these examples within the framework of the Tate-Schaferavich group \cite{Morrison:2014era} (or more generally, the group of CY torsors \cite{Bhardwaj:2015oru}). Within the following Sections we base our analysis of the particle spectrum on the smooth, genus one fibered manifolds themselves and assume that trivial uplifts from M-theory to F-theory exist (we will refer to this as the {\it``Working Assumption"} in later sections). However, this point is certainly deserving of more study and we will look at the physics of such uplifts more explicitly in a separate work \cite{Anderson19}. 

To construct quotients of elliptically fibered manifolds it is necessary to systematically understand how the discrete symmetry acts on the fibers/sections. To this end, we are aided by previous explorations such as \cite{Donagi:1999ez,Anderson:2018kwv,Bouchard:2007mf} which produce discrete symmetries of elliptic CY threefolds by demanding that the discrete action maps sections into one another in a fibration-preserving manner. As we will review in Section \ref{sec:quotients}, this can be accomplished in some cases by demanding that the fibers are of a form to support Mordell-Weil (MW) Torsion. In such a construction the rank of MW torsion in a covering space geometry is tied to the discrete symmetry action and hence to the order of the multiple fibers and non-trivial $\pi_1(\tilde{X})$ of the quotient manifold. 

In many cases we find that upstairs CY threefolds with non-trivial Abelian or non-Abelian gauge groups lead only to discrete gauge symmetries after quotienting. In particular we will demonstrate in the following sections that this quotient approach provides a powerful tool in explicitly building F-theory models with high rank discrete gauge groups (and for which no existing tuned Weierstrass models were previously known). 

The outline of this paper is as follows. In Section \ref{sec:quotients} we review the essential features of quotients of CY threefolds and the associated F-theory physics in $6$-dimensions. In particular, we review the physics associated to fixed points in the base of the fibration and the role played by multiple fibers in the genus-one fibration over these points. We also provide a systematic analysis of the matter spectrum associated to the downstairs theory in terms of the upstairs covering space theory. In Section \ref{sec:examples} we provide a number of concrete examples of quotient manifolds, including one leading to a $\mathbb{Z}_6$ discrete gauge group. In Section \ref{sec:bounds} we explore quotients of the so-called ``split bi-cubic'' or ``Schoen Threefold'' \cite{Candelas:1987kf,schoen} with Hodge numbers $(h^{1,1},h^{2,1})=(19,19)$ where a systematic classification of possible discrete quotients is known \cite{Bouchard:2007mf}. In this section we also collect observations from these quotient constructions to comment on possible bounds for discrete gauge symmetries appearing in $6$-dimensional F-theory compactifications (although a true bound is still an open question). In Section \ref{sec:conclusion} we conclude and discuss future directions. Some technical details are deferred to the Appendices.

\section{F-theory on quotient manifolds}
\label{sec:quotients}
In this section we discuss some of the general properties and constraints on the quotient geometries (and associated 6-dimensional F-theory physics) that will be considered throughout this work. Explicit examples are provided in Section~\ref{sec:examples}. 

\subsection{Covering geometries and their quotients}

Following \cite{Donagi:1999ez} we review the properties of smooth genus-one fibered threefolds $\tilde{X}$ with non-trivial fundamental group and their covering geometries $X$. We start by assuming that $X$ is a smooth, torus-fibered Calabi-Yau threefold over a smooth two fold base $B$. 
\begin{align}
\begin{array}{ccl}
\mathcal{T}& \rightarrow & X \\
&& \downarrow \pi\\
&&B
\end{array}
\end{align}

In the following we want to consider quotients\footnote{It should be noted that smooth quotients of (complex) CY n-folds by freely acting discrete automorphisms do not exist for $n$ even. For $K3$ surfaces and CY 4-folds, the quotienting process changes the anti-canonical class (e.g. the Enriques quotient of $K3$). As a result, the powerful relationship between the upstairs and downstairs theories studied here exists only in 6-dimensions.} of $X$ by free, cyclic and finite groups $\Gamma_{n}$ of order $n$ to obtain a new Calabi-Yau manifold, $\tilde{X}$. We place the important additional requirement on this group action that it preserves the fibration and as such, the quotient geometry can be used in a compactification of F-theory. In order for the quotient to be a Calabi-Yau manifold, $X$ must be equipped with a discrete automorphism $\Gamma_n \in \text{Aut}(X)$ that preserves its holomorphic three form. To ensure that the quotient preserves the fibration, we will choose the group action to be decomposable as
\begin{align} \label{fac}
g_{n} = g_{f} \circ g_b \;\;\; \forall \;\;  g_n \in \Gamma_n \, .
\end{align}
Here $g_f$ acts solely on fiber coordinates and $g_b$ acts solely on base coordinates in a given set of local trivializations which cover the base manifold. The $g_b$ will be elements of some group $\Gamma^B_m \subseteq \Gamma_n$ and the $g_f$ elements of some group $\Gamma^F_q \subseteq \Gamma_n$. In fact, in most of the explicit examples we will consider, the projection map takes a very simple form in terms of simply deleting some ambient space coordinates and the symmetry action on the Calabi-Yau 3-fold will descend from a linear action on the ambient space. In these cases we will have a similar factorization to (\ref{fac}) for the action on the ambient coordinates as well. In addition, all of the cases we will consider in later sections have $\Gamma^B_m = \Gamma_n$.

In a situation such as the one we have described in the previous paragraph, the base of the fibration associated to $\tilde{X}$ is
\begin{align}
\widetilde{B} = B/\Gamma_{m}^B \, .
\end{align}
We will require that the action of $\Gamma^B_{m}$ on $B$ admits at most fixed points under subgroups of $\Gamma^B_{m}$ of order $p$. These fixed points descend to singular points on $\widetilde{B}$ which are $A_{l-1}$ orbifold singularities with  $l=\text{gcd}(p,m)$. The resolution of each of these singular points would require $l-1$ exceptional curves of self intersection $-2$. Despite the fact that we will generically obtain a singular base upon quotienting, the full threefold $\tilde{X}$ can in fact remain smooth without blowing up the fixed points in the base if the fibers over those points are multiple. We will discuss this point in more detail in Section \ref{ssec:MultiFibers}. A superconformal matter sector is expected to be associated to each of these singular points \cite{DelZotto:2014hpa}. 

The action on the fiber of $\Gamma^F_q$ can be considered in more detail. In general, the covering geometry $X$ has singular fibers over the discriminant, $\Delta$, of the fibration which is co-dimension one in the base. For simplicity, we will require that the fixed points in the base miss this discriminant:
\begin{align}
S \cap \Delta = \varnothing \, .
 \end{align}
One way\footnote{But not the only way. See Section \ref{u1_eg} for an example of a different global fiber action.} to avoid fixed points in the total space $X$ can be accomplished by taking the actions $g_f \in \Gamma^F_q$ to be translations along the fiber. We will consider two types of fibrations admitting a group action which is a combination of an involution pulled back from the base and such a fiber-wise shift.

\begin{enumerate}
\item The fibration admits no section but only a multi-section $s^{(n)}$ of order $n$ such that $X$ is a genus-one fibration. The image of a point $b \in B$ under the multi-section
\begin{align}
\sigma^{(n)}(b) \in X 
\end{align}
is $n$ points on the associated torus fiber and $\Gamma^F_q$ acts as a translation that maps this set of solutions into one another.
\item The fibration admits a section $s_0$, the zero-section, giving the threefold $X$ the structure of an elliptic fibration. It is important that the zero-section not be invariant under $\Gamma_{q}^F$, as this would lead to fixed points in $X$ over the fixed points in the base. This implies the presence of $n$ additional sections $\sigma_n$ (or at least structure which echos such behavior over the $\Gamma_{m}^B$ fixed points). Note that, given the finite nature of the group $\Gamma_{q}^F$ this implies that the sections concerned should be torsional.
\end{enumerate}
The first case has been studied in detail in \cite{Anderson:2018heq} and so, in this work, we will focus mainly on the second of these two cases, first systematically used in string theory in \cite{Donagi:1999ez,Donagi:2000zf}. The action of the shift symmetry on the sections of the elliptic curve is precisely realized by the Mordell-Weil (MW) addition law \cite{Donagi:1999ez,Donagi:2000zf,Braun:2017feb}, denoted by $\oplus$, of rational sections, with the zero-section being its neutral element. As mentioned above, requiring finiteness of $\Gamma_{q}^F$ causes it to induce an action on the torsion part of the Mordell-Weil group \cite{MirandaTorsion,Morrison:2012ei}. After choosing a generating element $\sigma_1$ of the torsional sections, $\Gamma_{q}^F$ induces a translation $\hat{\Gamma}_{q}^F$ among the set of torsion sections as follows.
\begin{align}
\hat{\Gamma}_{q}^F:\, \sigma_i \xrightarrow{\oplus \sigma_1} \sigma_{i+1} \, .
 \end{align}
It is important to emphasize that this structure need not be realized globally. It is in fact only necessary that the fibration takes this form locally over the $\Gamma_{m}^B$ fixed points to guarantee a smooth quotient. Indeed, it can happen that a set of $n$ sections $\sigma_i$ mirror the form that torsional sections would take locally over $\Gamma_{m}^B$ fixed points such that a quotient is possible. We will return to this issue when we see examples of this phenomenon later on. 

\subsection*{Properties of the quotients and their implications for F-theory}

In this sub-section we will recall some geometrical properties of the quotient manifolds that we will be considering and discuss the implications of these for the F-theoretical physics that arises. First, we recall that indices, such as the Euler number, get divided by the order of the quotient in passing to the downstairs space.
\begin{align}
\chi(\tilde{X}) = \frac1n \chi(X) \, .
\end{align}

In fact, we will need a somewhat more refined understanding of what happens to $h^{1,1}$ and $h^{2,1}$ of the manifold under quotienting. For this we recall that, if the upstairs manifold is elliptically fibered, we can, by the theorem of Shioda, Tate and Wazir \cite{shioda,shioda2,COM:213767}, identify divisors as being either vertical or horizontal in nature. More precisely, we have the following division,
\begin{eqnarray}
h^{1,1}(X)=h^{1,1}(B) +h^{1,1}_h(X) +h^{1,1}_f(X) \;,
\end{eqnarray}
were $h^{1,1}_h(X)$ and $h^{1,1}_f(X)$ are the number of independent sections that generate the free Mordell-Weil group and the number of fibral divisors, respectively (and we assume flat fibrations).

Given that the divisors of the quotient manifold $\tilde{X}$ descend from a subset of the divisors of the covering space, $X$, we will also be able, in the examples we study, to use their antecedents on the covering space to classify the divisors on the quotient into horizontal and fibral types as well (note that in the following we will be employing the {\it ``Working Assumption"} about the M-/F-theory uplift discussed in Section \ref{sec:intro}):
\begin{align}
h^{1,1}(\tilde{X}) = h^{1,1}(\tilde{B}) +  h^{1,1}_h(\tilde{X}) + h^{1,1}_f(\tilde{X}) \, ,
 \end{align} 
Upon embedding such a geometry in F-theory, one can, in the examples we will consider, then read off some of the physical features of the resulting theory from pieces of this decomposition. In particular, we will use that
\begin{align}
h^{1,1}(\tilde{B})-1= T_{(1,0)}  \, ,  \qquad h^{1,1}_f(\tilde{X})-1 = \text{rk}(G) \, ,
\end{align}
where $T_{(1,0)}$ is the number of tensor multiplets and $\text{rk}(G)$ is the rank of the non-abelian gauge group.

The quantities $h^{1,1}(X)$ and $h^{1,1}(\tilde{X})$ can be different or the same, depending upon the nature of the action of $\Gamma_n$. Defining $\Delta h^{1,1}_b = h^{1,1}(\tilde{B}) - h^{1,1}(B)$ and $\Delta h^{1,1}_f = h^{1,1}(\tilde{X})-h^{1,1}(X)$ we then have a number of possibilities as to the situation that could occur on quotienting:
\begin{enumerate}
%\item $\Delta h^{1,1}_b=\Delta h^{1,1}_f=0$: Gauge symmetry and number of $(1,0)$ tensors is unchanged.
\item $\Delta h^{1,1}=0$: Gauge symmetry and number of $(1,0)$ tensors is unchanged.
\item $\Delta h^{1,1}_b <0$: Number of (1,0) tensors reduced in the quotient.
\item $\Delta h^{1,1}_f <0$: Rank of non-Abelian gauge symmetry reduced in the quotient.
\item $\Delta h^{1,1}_h <0$: Rank of Abelian gauge symmetry reduced in the quotient.
\end{enumerate}
Cases $2-4$ could, of course, happen in the same geometry. In the following we want to show that quotients of elliptic fibrations necessarily have feature 3 or 4. 

As argued above, if $X$ is elliptic we need a set of additional sections $\sigma_i\, ,  i=1\ldots n-1$ to be related by the translations along the fiber over the fixed points.
Since each section intersects the fiber $\mathcal{E}$ once it follows that the resulting geometry is a genus-one geometry with n-sections
\begin{align}
\sigma_i (X) \cdot \mathcal{E} = 1 \, ,\quad    \sigma_i(\tilde{X}) \sim \sigma_0 \, \forall i\, , \\
\sigma_i \rightarrow \sigma^{(n)}\, , \qquad\sigma^{(n)} \cdot \mathcal{E} = n \, . 
\end{align}
A central object in our story will be the Shioda map $\Sigma(\sigma_i)$, that is a group homomorphism of a section $\sigma_i$ into  $h^{1,1}(X)$ of the threefold.
If these sections live in the free part of the MW group then each one of them corresponds to a linearly independent divisor leading to an $U(1)^{n-1}$ gauge symmetry in the 6-dimensional F-theory \cite{Park:2011ji}. The associated divisor is given as the image of the section under the Shioda map.
\begin{eqnarray} \label{shioda}
\Sigma(\sigma_i) = [\sigma_i]-[\sigma_{0}] - \pi^* (D_b) + K_{i,m} (C^{-1})^{m,n}  D_{f,n}    \, ,
\end{eqnarray}

Here, $[\sigma]$ is the divisor class of the zero set of the section $\sigma$ and $K_{i,m}= ([\sigma_i]-[\sigma_0])\cdot{\cal C}_m$ where ${\cal C}_m$ is the fiber $\mathbb{P}^1$ of the exceptional fibral divisor $D_{f,m}$. The object $C^{-1}$ is the inverse of the Cartan matrix of the gauge algebra associated to the fibration which is given by $C_{mn}=-D_{f,m}\cdot {\cal C}_n$ in terms of intersections. Finally $D_b$ is some divisor in the base which is chosen in order to ensure that $\Sigma(\sigma_i)$ has zero intersection with any divisor which is vertical or the zero section. The presence of such a correction term signals the presence of a non-abelian gauge group, which in the cases we will study will be a non-simply connected gauge group of the form $(G \times U(1)^{n-1})/\mathbb{Z}_t$ \cite{Cvetic:2017epq,Grimm:2015wda}.

If sections $\sigma_i$ in the free part of the Mordell-Weil group get identified under a quotient, such that $[\sigma_i] \sim [\sigma_0]$ for  $i=1\ldots n-1$ for example, then the zero-section becomes an n-section and the descendants of all the Shioda maps $\Sigma(\sigma_i)$ trivialize in the quotient. The $U(1)^{n-1}$ gauge group factor does not appear in the quotient geometry due to monodromies and is replaced instead with a $\mathbb{Z}_n$ symmetry. The discrete charges of the matter in the resulting theory can be understood in terms of the $U(1)^{n-1}$ charges of the upstairs model. As always in a valid F-theory compactification, the change in the degrees of freedom that can be computed geometrically ensures general anomaly freedom, as we show in this case in Subsection~\ref{sec:Anomalies}.
  \begin{figure}[t!]
 \begin{picture}(0,150)
	  \put(0,0){ \includegraphics[scale=1.7]{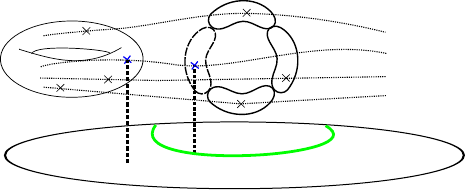}}
 \put(220,30){{\LARGE $\xrightarrow{\Gamma_n}$}}
  \put(240,0){ \includegraphics[scale=1.7]{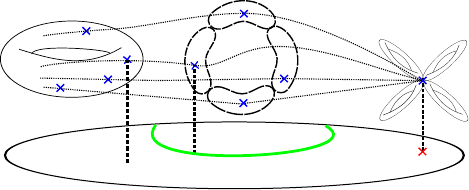}}
  \end{picture}
  \caption{\label{fig:torsionquotient}{\it A torsion model on the left with an $SU(n)/\mathbb{Z}_n$ gauge group and the covering space description of its free quotient on the right. The torsion sections fuse into a multi-section with the additional effect to identify all $SU(n)$ resolution divisor with the affine node.}}
  \end{figure}
  
  We can also obtain sections $\sigma_i$ in the upstairs geometry that are global torsion. Similarly to the case of free sections, there exists a torsion Shioda map upstairs \cite{Mayrhofer:2014opa} whose image is a trivial divisor and therefore does not contribute to $h^{1,1}(X)$. However the effect of the torsional section is subtle: it leads to a singular Weierstrass model associated to a gauge algebra $\mathcal{G}$ with a non-trivial $\mathbb{Z}_n$ center \cite{Aspinwall:1998xj,Mayrhofer:2014opa}. The mere existence of torsional sections, then, implies that a non-abelian gauge group, associated to non-trivial vertical divisors $D_j$ in the resolved geometry, will be non-simply connected. The effect of the quotient can also nicely be seen in the structure of the Coulomb chambers in the 5-dimensional M-theory \cite{Esole:2017hlw} which is coarser when the torsion is present.

The interplay of resolution divisors with the torsional sections is non-trivial, precisely because the torsional object encodes the non-simply connectedness of the upstairs gauge group, as we show explicitly in examples in Section~\ref{sec:examples}. The key observation \cite{Mayrhofer:2014opa} is, that the torsional section $\sigma_i$ intersects the resolution divisors $D_{f,n}$ in a non-trivial fashion $\sigma_i \cdot D_{f,n}  = K_{i,n}$  such that the torsional Shioda map assigns to every torsional section a divisor 
 \begin{align}
 \Sigma(\sigma_i) =  [\sigma_i] - [\sigma_0] - \pi^* (D_b) +  K_{i,m} (C^{-1})^{m,n} D_{f,n}  \, , 
 \end{align}
with $C^{-1}$ being the inverse Cartan matrix leading to fractional coefficients in the above expression. As $\Sigma(\sigma_i)$ is trivial, we may write
\begin{align}
 \Xi_n := [\sigma_i] - [\sigma_0 ]- \pi^* (D_b) = -K_{i,m} (C^{-1})^{m,n} D_{f,n}  \, .
\end{align}
This can then be interpreted as a n-torsional element of the cohomology $H^{1,1}(X,\mathbb{Z})/\langle [D_{f,n}] \rangle$:
\begin{align}
n \cdot \Xi_n = \sum a_m D_{f,m} = 0  \text{ mod } [D_{f,m}]  \, .
\end{align}

Having reviewed the role of MW torsion in the description of non-simply connected gauge groups in the F-theory we are now in position to take the quotient. If a quotient identifies a set of sections $\sigma_i \sim \sigma_0$ inducing 
\begin{align}
[\sigma_i]-[\sigma_0] = 0 = \pi^*(D_b) -K_{i,m} (C^{-1})^{m,n} D^{\mathcal{G}}_n \quad  \forall i \, ,
\end{align}
then this adds a linear equivalence relation among the resolution divisors and $\pi^*(D_b)$ for each $\sigma_i$ so identified. In other words, as the torsional sections $\sigma_i$ get identified with the zero-section, then resolution divisors of $\mathcal{G}$ that are intersected by those divisors get removed as independent divisor classes by being related to $\pi^*(D_b)$ as shown in Figure~\ref{fig:torsionquotient}. Note that in the case when a $U(1)$ generating section intersects some resolution divisor  an analogous effect occurs, where non-abelian gauge group factors get reduced as compared to the covering space upon the identification of resolution divisors with the affine node.

Naturally, the above discussion of divisor classes has important consequences for gauge symmetry and matter content of the 6-dimensional effective F-theory description. We can see this by taking the F-theory limit from M-theory and by recalling the origin of vector and hypermultiplets of some ADE resolved singularities over some genus $g$ curve in the base \cite{Witten:1996qb}.
In the covering elliptic fibration the affine $\mathbb{P}^1$ stays at finite size, when taking the F-theory limit and is identified by the intersection with the zero-section. Counting all fibral curves $C$ with self intersection $-2$ that are shrinkable and hence do not contain the affine $\mathbb{P}^1$ as a component, leads to a vector and $g$ hypermultiplets and yields the adjoint representation. On the other hand all curves of self intersection $-2$ that do contain the affine $\mathbb{P}^1$ and can not be shrunken contribute $g-1$ hypermultiplets again comprising the full adjoint representation. 

How does the situation change in the quotient theory? We have already seen that the quotient reduces the number of sections. In a situation where all of the generating sections are identified with the zero-section in the quotient, none of the resolution divisors become shrinkable and therefore do not contribute vector multiplets to the downstairs theory. However there is still the contribution of the non-shrinkable curves that give the same count as the adjoint representation in the covering geometry. Thus we find the following amount of additional discrete charged hypermultiplets from the adjoint representation of the covering geometry
\begin{align}
\widehat{H}'_{discrete} = \text{rt}(\mathcal{G}) \frac{(g-1)}{n} \, ,
\end{align}
where we denote $\text{rt}$ as the charge dimension of the adjoint group $G$. Also note the additional reduction by $n$ due to the reduction of intersection numbers, that we will explain in the following sections in more detail.

In total we can interpret the residual gauge group as the one of the covering theory, fully broken by monodromy. These effects capture an important part of the perturbative degrees of freedom that are necessary to prove general anomaly cancellation in Subsection~\ref{sec:Anomalies}. However this is only enough to prove the anomalies of the gauge sector but not the gravitational ones, as those are also sensitive to the fixed points with the multiple fibers, which we discuss below. 

\subsection{Multiple fibers and hyperconifolds}
\label{ssec:MultiFibers} 
In this section we review the phenomenon of multiple fibers in the simple example of a rational elliptic surface $S$, following the discussion in\cite{Griffiths}. The surface $S$ admits a holomorphic map that projects to the complex one dimensional base
\begin{align}
\pi: \, S \rightarrow B_1 \, .
\end{align}
At a generic point $b \in B_1$, the pullback of a local coordinate that vanishes at $b$, that is $\pi^* z$ vanishes to order $n=1$ along the fiber $\pi^{-1}(b)$. If, over a special point $b_0 \in B_1$ this vanishing is instead of order $n>1$, the fiber over $b$ is said to be multiple of order $n$.

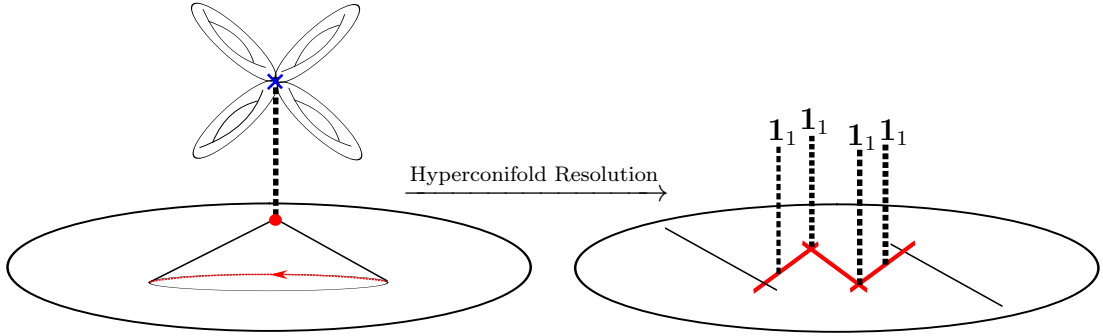
\begin{figure}[t!]
\begin{center}
\begin{picture}(0,110)
\put(-200,0){\includegraphics[scale=2.5]{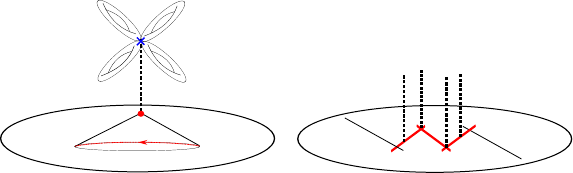}}
\put(-50,50){$\xrightarrow{\text{Hyperconifold Resolution}}$}
\put(88,73){$\mathbf{1}_1$} 
\put(100,77){$\mathbf{1}_1$}
\put(130,73){$\mathbf{1}_1$}
\put(118,71){$\mathbf{1}_1$}
\end{picture}
\caption{\label{fig:hyperconifold}{\it Depiction of a hyperconifold resolution of an n-multiple fiber corresponding to the tensor branch of the $\mathcal{A}_n$ discrete gauged superconformal matter with additional discrete charged states \cite{Anderson:2018heq}.}}
\end{center}
\end{figure}

There is a textbook construction of multiple fibers that mirror what we will see in case of a compact Calabi-Yau threefolds. First pick $B_1$ to be a local neighborhood of the point where the multiple fiber will be located, with coordinate $z$. We also take an elliptic curve  $\mathcal{E}$ with associated coordinate $\omega$ and complex structure $\tau$. Finally, we pick an order $n$ automorphism
\begin{align}
\phi_n : \quad \mathcal{E} \times B_1 \rightarrow \mathcal{E} \times B_1  \, ,
\end{align}
 acting as a free quotient on the total space of the direct product of  $\mathcal{E}$ and $B_1$ but as an orbifold on the base and a translation in the fiber,
 \begin{align}
 \phi_n (\omega, z) = (\omega+ \frac{\tau}{n} , e^{2\pi i/n} z ) \, .
 \end{align}
Denoting the quotient surface as
\begin{align}
 S =  (\mathcal{E} \times B_1) / \phi_n \, ,
\end{align}
then the induced morphism $\widehat{\pi}$ coming from the following map on the covering geometry, 
 \begin{align}
\pi (\omega, z) \rightarrow z^n \, ,
 \end{align}
is well defined on $S$ and forms a suitable projection. Labeling a coordinate on the base of the quotient by $\lambda=z^n$, we can then infer the structure of the fibers from the covering space. Picking a generic point $\lambda \neq 0$, the pullback on the covering space $\phi_n^* \widehat{\pi}^{-1}(\lambda)$   consists of the n curves on the covering geometry at points $z_b$ with $z_b^n = \lambda$. However over $\lambda=0$ there exists only a single  elliptic curve on the covering geometry. Thus $\widehat{\pi}^{-1}(0)$ is an order $n$ multiple fiber.\\\\
The quotient constructions that were used in \cite{Anderson:2018heq}, as well as in this work, are structurally very similar to the above but with the base being a compact twofold embedded into a smooth threefold. We note, that in F-theory the base, $B$, is the physical compactification space and the symmetry by which we quotient, $\Gamma_n$, acts like a standard orbifold. This introduces a non-standard $\mathcal{A}_n$ (2,0) superconformal matter sector into the low energy theory that naively contributes to anomalies in the same manner as a $(1,0)$ tensor and a neutral hypermultiplet
\begin{align}
\label{eq:20Tensors}
\mathcal{A}_n \sim (n-1) \times ( T_{(1,0)} \oplus H_{\mathbf{1}_0}) \, .
\end{align}
However, as was shown in \cite{Anderson:2018heq} these sectors differ in a striking manner from the standard (2,0) superconformal matter exactly due to the presence of the multiple fibers. They are gauged under a $\mathbb{Z}_n$ discrete symmetry. This gauging is visible when going to the tensor branch of the theory and is nicely related to the resolution of a hyperconifold transition utilizing a Lens space \cite{Davies:2009ub,Davies:2011is,Davies:2013pna}. Such a transition, resolving the fixed point in the base and removing the multiple fiber, is characterized by the following change of Hodge numbers.
\begin{align} \label{tho}
\Delta (h^{1,1},h^{2,1})=(n-1,-1) \, 
\end{align}
Over the exceptional divisors on the resolution side of the hyperconifold, n fibers of $I_2$ type are found at codimension $2$ in the base, giving rise to discrete charged singlets\footnote{In the associated Weierstrass model of the Jacobian fibration, the fiber singularities are in fact terminal, but they are smooth in the genus-one fibration \cite{Arras:2016evy,Grassi:2018rva}.} Since, from (\ref{tho}) we loose a neutral hyper-multiplet in the complex structure sector during this transition, we find that the matter localized at the orbifold fixed point and its subsequent resolution that contributes to the anomaly should be thought of not as in (\ref{eq:20Tensors}), but rather, after the hyperconifold transition, as follows.
\begin{align}
\mathcal{A}_n \oplus H_{\mathbf{1}_0} \xrightarrow{\text{Hyperconifold}} (n-1) \times T_{(1,0)} \oplus n \times H_{\mathbf{1}_1} \, .
\end{align}
In summary, the gauging reflects itself in the resolved geometry by the additional discrete charged singlets, as opposed from the naive decomposition Eq.~\eqref{eq:20Tensors} as depicted in Figure~\ref{fig:hyperconifold}.
\\
 We will not discuss the tensor branch of these theories further and treat them, when it comes to anomalies, simply as (2,0) superconformal matter keeping in mind that they are actually discrete charged and hence are seen to actually be (1,0) theories.

\subsection{6-dimensional spectrum and anomaly cancellation}
\label{sec:Anomalies}

Given that the geometries $\tilde{X}$ are genus one fibered Calabi-Yau manifolds, we expect them to be associated to good 6-dimensional F-theory compactifications with all anomalies canceled. In addition, we construct these quotients in such a manner that we expect that the massless degrees of freedom  descend in a well defined fashion from those of the theory associated to the covering space. We can check this hypothesis by verifying that the anomalies are still canceled. That this is indeed the case for the different types of quotients we will consider is what we will show in the following. Here we talk in some generality before proceeding to some explicit examples in Section~\ref{sec:examples}.

\vspace{0.1cm}

Starting with a covering geometry $X$ over a smooth Fano base $B$, we take a free finite $\mathbb{Z}_n$ quotient reducing the fundamental domain of $\tilde{B}$ by $\frac1n$. However as the action on the base is generically non-free we obtain, in the examples we will consider, codimension two fixed points with additional superconformal matter.
  In the following we use that all divisors $D\in \{ b, b_{mn}, a \}$ in $\widehat{B}$ associated to gauge divisors, U(1) height pairings and, the canonical class of the base respectively are Cartier and hence do not intersect the above mentioned fixed points. Therefore, we expect the (2,0) superconformal matter that appears on these singularities to be gauged only under the discrete symmetry \cite{Anderson:2018heq}. For the ADE divisors $b$, this in fact follows from the requirement of a smooth quotient action.  
   In addition to those states, the quotient affects the 6-dimensional SUGRA by reducing the number of tensors $T$, the number of Abelian symmetries and by giving rise to a  smaller non-Abelian gauge group in general. We will call the commutant of the downstairs gauge group inside the upstairs one $G^\prime$. 
  The change in the hypermultiplet sector can then be obtained from the covering geometry simply by using that the reduction of the fundamental domain of the base by a factor of $1/n$ leads to the same reduction in the number of the charged hypermultiplets. This reduction is readily obtained from the intersection numbers in the quotient base and the fact, that the hypermultiplets are never localized over the fixed points, by construction. Note that in assuming that the intersection numbers in the base divide by the order of the discrete group we are assuming that the integral basis of divisors on $\tilde{B}$ descends directly from that on $B$ (note that for some discrete actions a change of basis is required, but we will not consider such examples in the present work, see e.g. \cite{Anderson:2009mh,Braun:2009mb} for examples of such basis issues).
  
  We summarize the change in the full hypermultiplet sector as
\begin{align}
\label{eq:Hyperchange}
\begin{split}
\widetilde{H}_{\text{charged}} =&\frac1n H_{\text{charged}} \, , \\
\widetilde{H}_{\text{adjoint}}(G) =&\frac1n H_{\text{adjoint}}(G)+\frac{(n-1)}{n}\left(V(G)-\text{rk}(G)\right)\, , \\
\widetilde{H}_{\text{neut}}=& H_{\text{neut}} - \Delta h^{1,1} +\frac12 \chi \frac{(n-1)}{n} \, ,\\
\widetilde{H}_{discrete}^\prime =& \frac1n H_{\text{adjoint}}(G^\prime) -\frac1n (V^\prime(G^\prime)-rk(G^\prime))\, .
\end{split}
\end{align}
As discussed above, the change in $H^{1,1}$ can either be zero or have three possible contributions $\Delta h^{1,1} = \{0,\Delta T,   \text{rk}(\text{MW}),  \text{rk}(G^\prime ) \}$. Clearly it is a requirement that all the above multiplicities are integer valued giving non-trivial constraints on the covering geometry (that are expected to hold in the presence of a free order $n$ automorphism $\Gamma_n$). 

In the following we show that the above spectrum indeed captures all massless degrees of freedom such that all anomalies in the quotient theories are canceled. The anomalies of the unbroken gauge groups in the quotient theory are easily checked using the above change in the spectrum. The intersections on the quotient base $\tilde{B}$ simply get multiplied by $\frac1n$ as long as the divisors involved are Cartier, as is guaranteed by smoothness of the quotient. For more details see \cite{Anderson:2018heq}. The mixed gauge-gravitational anomaly 
\begin{align}
\textnormal{Gauge}^2 \cdot \textnormal{Grav}^2: \qquad   -\frac13 \left( (1-H_{\text{adj}}) A_{adj}+ \sum_\mathbf{R} A_\mathbf{R} H_\mathbf{R} \right) = b \cdot a \, ,
\end{align}
is simply divided on both sides, on the left due to the reduced spectrum, and on the right due to the divided intersections of the divisors $a$ and $b$ on the base. 
   Hence only the gravitational anomalies need a more careful inspection. 
   
The anomaly cancellation condition
\begin{align}
9-T = (K_b^{-1})^2 \, ,
\end{align}
which is satisfied in the upstairs theory, reduces to
\begin{align}
 T_{(2,0)}-  \Delta T  = \frac{(n-1)}{n} (K_b^{-1})^2 \, .
\end{align}
The number of (2,0) tensors contributed by each fixed point is simply given, in terms of its order $l$, by $l-1$. 

Finally we turn again to the irreducible anomaly
\begin{align}
 H_{\text{neut}} + H_\text{adjoint}+H_{\text{charged}}-V+29T -273 = 0 \, ,
\end{align}
We can use this to rewrite the the Euler number as  
\begin{align} \label{eulerguy}
\frac{(n-1)}{2n}\chi =&  \frac{(n-1)}{ n} \left( \text{rk}(G)+T+3-H_n            \right)\, ,
 \nonumber
 \\
=& \frac{(n-1)}{ n} \left(H_{\text{charged}}+ H_{\text{adjoint}}+\text{rk}(G)-V(G) +30 T-270     \right)\, ,
 \nonumber \\
 =&\frac{(n-1)}{ n} \left(H_{\text{charged}}+ H_{\text{adjoint}}+\text{rk}(G)-V(G) \right) - 30(T_{(2,0)}- \Delta T) \, .
\end{align}
The above equation can be used to deduce cancellation of the gravitational anomaly in the quotient theory which admits the reduced spectrum
\begin{align}
\label{eq:downstairsGrav}
\widetilde{H}_{neut} + \widetilde{H}_{charged}+\widetilde{H}_{adjoint}+\widetilde{H}_{charged}^\prime-(V-V(G^\prime  ))+29(T-\Delta T)+30 T_{(2,0)} -273 =0 \, .
\end{align}
 In the following we check anomaly cancellation for several cases individually for clarity. 
 
 \subsubsection*{Tensor Reducing Quotients}
We start with tensor reducing quotients where the change in Hodge number can be fully identified with the changed number of tensors, $\Delta h^{1,1} = \Delta T$, and hence there is no change in the total gauge group\footnote{This implies a genus-one fibration on the covering geometry and a discrete symmetry already present there.}. To be fully concrete, the full charged matter spectrum in the quotient theory is now reduced to 
\begin{align} 
\begin{split}
\widetilde{H}_{\text{charged}}=&\frac1n H_{\text{charged}}\, , \\
\widetilde{H}_{\text{adjoint}} =&\frac1n H_{\text{adjoint}}-\frac{(n-1)}{n}\text{rk}(G) \, , \\
\widetilde{H}_{\text{neut}}=& H_{\text{neut}} - \Delta T +\frac12 \chi \frac{(n-1)}{n} \, , 
\end{split}
\end{align}
in addition to the new discrete charged (2,0) strongly coupled sector.
Plugging in the change in hypermultiplets in Eq. \eqref{eq:Hyperchange} the gravitational anomaly of the quotient theory becomes 
\begin{align*}
&\widetilde{H}_{neut} + \widetilde{H}_{charged}+\widetilde{H}_{adjoint} -V +29(T-\Delta T)+30 T_{(2,0)} -273 \, , \\
&= H_{neut} - \Delta T+ \frac{(n-1)}{2n} \chi +\frac1n H_\text{charged}+\frac{1}{n}H_{\text{adjoint}}+\frac{(n-1)}{n} \text{rk}(G)\\ &\qquad -V+29(T-\Delta T)+30 T_{(2,0)}-273 \, , \\
&=0 \, ,
\end{align*}
and hence is also satisfied upon using (\ref{eulerguy}). 
\subsubsection*{Mordell-Weil reducing quotients}
We next consider a case where the change in the MW rank, and thus the number of $U(1)$'s, accounts entirely for the change in Hodge number $\Delta h^{1,1} =   \textnormal{rk}( MW)$. The spectrum gets reduced as in the case before, with the exception that Abelian charges are now interpreted as discrete ones, such that the charged hypers get reduced to
\begin{align}
\widetilde{H}_{discrete} =&\frac1n H_{\text{charged}}  \, .
\end{align} 
Note from above, that we also have to include the Abelian charged singlets, that are now discrete charged ones.
With this change the gravitational anomaly in the quotient theory is satisfied as well
\begin{align}
&\widetilde{H}_{neut} + \widetilde{H}_{charged}+\widetilde{H}_{adjoint}(G)+\widetilde{H}_{discrete} -V(G) +29 T+30 T_{(2,0)} -273 = 0 \, .
\end{align}

\subsubsection*{Non-Abelian Group reducing quotients}
Finally we consider the case where $\Delta T = 0$ and we can identify the change in Hodge number entirely with the reduction of the resolution divisors of the gauge group, as determined by $G^\prime$. In this situation, in the quotient theory, we simply have a gauge group $G$ with $T_{(2,0)}$ superconformal tensors and a reduced amount of hypermultiplets to solve the gravitational anomaly
\begin{align}
\label{eq:grav5}
&\widetilde{H}_{neut} + \widetilde{H}_{charged}+\widetilde{H}_{adjoint}+\widetilde{H^\prime}_{discrete} -V(G) +29 T+30 T_{(2,0)} -273 = 0 \, .
\end{align}
This is indeed the case, using the following charged hypermultiplet spectrum
\begin{align}
\widetilde{H}_{\text{charged}}=& \frac1n H_\text{charged} \, \\
\widetilde{H}_{\text{adjoint}}(G)=& \frac1n H_\text{adjoint}(G)+\frac{n-1}{n}(V(G)-\text{rk}(G)) \\
\widetilde{H^\prime}_{\text{discrete}}=& \frac1n H_\text{adjoint}(G^\prime)-\frac1n(V(G^\prime)-\text{rk}(G^\prime)) 
  \, 
\end{align}
  and neutral degrees of freedom
  \begin{align}
  \widetilde{H}_{\text{neut}}=&H_{\text{neut}}-\text{rk}(G^\prime)-30T_{(2,0)}\nonumber \\ &+\frac{n-1}{n}(H_\text{charged}+H_\text{adjoint}(G)+( \text{rk}(G)-V(G)))\nonumber \\
  &\, \, \,    \qquad \qquad +\frac{n-1}{n}( H_\text{adjoint}(G^\prime)+( \text{rk}(G^\prime)-V(G^\prime))           \, .
  \end{align}
 Using the gravitational anomaly for the neutral hypers of the covering theory, given as
 \begin{align}
 H_{\text{neut}}= V(G)+V(G^\prime)-29 T - H_\text{charged}-H_\text{adjoint}(G)-H_\text{adjoint}(G^\prime)+273 \, .
 \end{align}
one can then verify that \eqref{eq:grav5} is indeed satisfied and the gravitational anomaly is also satisfied in the quotient with this matter content. 

\section{Examples of quotient geometries}
\label{sec:examples}
Below we illustrate explicitly some of the possible quotient actions on elliptically or genus-one fibered CY 3-fold geometries. It will be demonstrated in each case that the F-theory physics of the theory associated to $\tilde{X}=X/\Gamma$ can be readily determined from the covering geometry, $X$. In something of an abuse of nomenclature we will refer to the effective 6-dimensional theory obtained by F-theory compactified on $X$ as the ``upstairs theory,'' while that associated to a compactification on $\tilde{X}$ will be referred to as the ``downstairs theory.''  As mentioned in Section \ref{sec:quotients}, it is important to recall that the upstairs and downstairs theories are not dynamically related in any way (and correspond to topologically very distinct geometries). However, it is a unique feature of the downstairs quotient geometries that they can be entirely specified in terms of $\Gamma$-invariant quantities in the upstairs geometry. In our context, this will allow us to describe the downstairs theories, which, as discussed in Section \ref{sec:quotients}, can have a multitude of complicated geometric features, in terms of their much simpler covering spaces.

In the case of elliptic fibrations, as mentioned in Section \ref{sec:quotients}, due to the theorem of Shioda-Tate-Wazir, it is clear that we can characterize the action of the discrete symmetry on divisors by whether the classes of horizontal and/or vertical divisors are reduced or preserved under the symmetry action. In the case that $h^{1,1}(\tilde{X}) < h^{1,1}(X)$, this will lead to the classes of examples outlined in Section \ref{sec:quotients}.

In the following sub-sections, we consider explicit CY quotients that illustrate each effect in isolation. We conclude this section with a more complicated example of a higher order quotient with non-trivial subgroups that both reduces the rank of the gauge group and reduces the number of tensors in order to set the stage for more general and complicated possibilities. The latter will be illustrated via the well-known Schoen manifold (with Hodge numbers $(h^{1,1},h^{2,1})=(19,19)$) in Section~\ref{sec:bounds}. 

\subsection{ Tensor reducing  $\mathbb{Z}_2$ quotient}
\subsubsection{The geometry}
The simplest class of examples to consider is one in which the discrete symmetry action identifies divisors in the base, $B$, of the fibration $\pi: X \to B$. In such a fibration, it is actually unimportant whether or not a section exists, since these base divisors, in either the elliptic or genus-one fibered case, play a clear role both geometrically and in the counting of tensor multiplets in the 6-dimensional theory. 

With this in mind, we begin with a quotient action that acts non-trivially on the base, $B$, of the genus-one fibration $\pi: X \to B$ and in particular, reduces the dimension of $h^{1,1}(B/\Gamma)$ compared to $h^{1,1}(B)$, so that the number of tensor multiplets in the downstairs theory is less than that of the upstairs theory.

Consider a simple direct product manifold as the ambient variety, ${\cal A}=\mathbb{F}_0 \times dP_3$, whose anti-canonical hypersurface will define the upstairs (i.e. covering) CY 3-fold, $X$. This ambient space can be torically realized as being associated to the convex hull of the polytope
\begin{align}
\label{eq:polyexample2}
\begin{array}{|cccc|cccccc|}\hline
x_0 & x_1 & y_0 & y_1 & z_1 & z_2 & z_3 & z_4 & z_5 & z_6  \\ \hline
1    & -1     & 0    & 0    & 0     &    0 &    0 &   0  &   0  &   0   \\
0    &   0    &  1    & -1  &  0    &    0 &   0  &  0   &  0   &  0  \\
0    &   0    &  0    &  0  &  -1   &    0  &  1  &  1  & 0    &  -1 \\
0    &  0    &   0    &   0 &  -1   &    -1 &  0  & 1  &  1   & 0  \\ \hline 
\end{array}_{(h^{1,1}=6,h^{2,1}=54)}^{(\chi= -96) } \, .
\end{align} 
(where the superscript denotes the Euler character and the subscript the Hodge numbers of the resulting CY threefold) resulting in the $\lambda_i \in \mathbb{C}^*$ equivalences
\begin{align}
\begin{split}
\label{eq:scaling}
dP_3&: (  z_1,z_2,z_3,z_4,z_5,z_6) \sim (\lambda_4  \lambda_1 z_1,\lambda_2   z_2,\lambda_3 \lambda_4 z_3,\lambda_1 z_4,\lambda_2 \lambda_4 z_5,\lambda_3 z_6 ) \, , \\
\mathbb{F}_0&: (x_0,x_1,y_0,y_1) \sim (\lambda_5 x_0, \lambda_5 x_1,\lambda_6 y_0,\lambda_6 y_1) \,  
\end{split}
 \end{align} 
and the Stanley-Reisner ideal (SRI)
\begin{align}
SRI: \{  x_0 x_1, y_0 y_1; z_1 z_3, z_1 z_4, , z_2 z_4, z_2 z_5, z_2 z_6, z_3 z_5,z_3 z_6, z_4 z_6 \} \, .
\end{align} 

To build a quotient threefold $\tilde{X}=X/\Gamma$, we fix a $\mathbb{Z}_2$ discrete symmetry 
acting on the ambient space as an orbifold on the $\mathbb{F}_0$ component and a 180$^\circ$ rotation on the $dP_3$ toric diagram acting on the coordinates as \cite{Braun:2017feb}
 \begin{align}
\label{eq:qutientEx2}
\Gamma_2:\, (x_j,y_j,z_i) \to ((-1)^j x_j ,(-1)^j y_j, z_{i+3}) \text{ for } j=1,2 \qquad i=1..6  \, .
\end{align}
The full ambient space admits $4\times4$ fixed points. Those of $\mathbb{F}_0$ lie over the intersection of toric divisors whereas those for dP$_3$ do not, due to the form of the SRI. Here, the fixed points satisfy the equation
 \begin{align}
 z_i = \lambda_i z_{i+3} = r_i\, ,
 \end{align}
 with $\lambda_i, r_i \in \mathbb{C}^*$. Using the $\lambda_4$ relation, these can be fixed to the following set of fixed points  
 \begin{align}
 fp_{\mathbb{F}_0/\mathbb{Z}_2}&: [x_0,x_1; y_0, y_1] = (\underline{0,1},\underline{0,1})\, ,\\
 fp_{dP_3/\mathbb{Z}_2}&:(z_1,z_2,z_3,z_4,z_5,z_6)=\left( \begin{array}{cc} +1,+1,+1&,+1,+1,+1 \\
 \underline{+1,-1,-1}&,+1,+1,+1 \end{array} \right)\, .
 \end{align}
 where different permutations are denoted via an underline.

In this example, we can view $B=dP_3$ as the base of the fibration and the genus-one fiber as a biquadric in $\mathbb{F}_0$ with hypersurface equation
\begin{eqnarray}
 p&=&(s_1^{(+)} y_0^2 + s_2^{(-)} y_1 y_0 + s_3^{(+)} y_1^2) x_0^2 + (s_5^{(-)} y_0^2 + 
    s_6^{(+)} y_1 y_0 + s_7^{(-)} y_1^2) x_0 x_1 \\ \nonumber&&+ (s_8^{(+)} y_0^2 + 
    s_9^{(-)} y_1 y_0 + s_{10}^{(+)} y_1^2) x_1^2
\end{eqnarray}
The functions $s_i$ are generic sections of $\mathcal{O}(K_{dP_3}^{-1})$. Once the discrete group action is imposed however, we must require the complete defining equation to be equivariant. Taking into account the $\mathbb{Z}_2$ action on the fiber, coefficient functions $s_i^{\pm}$ must transform equivariantly with $\pm$ eigenvalues. This equivariance requirement forces a tuning of the complex structure to yield the following
\begin{align}
s_i^{(+)}=& z_1^2 z_2 z_6 (z_5 z_6 a_{[1,i]}+ z_2 z_3 a_{[2,i]}) + 
 z_3 z_4^2 z_5 (z_2 z_3 a_{[1,i]} + z_5 z_6 a_{[2,i]}) \\ & + 
 z_1 z_4 (z_2^2 z_3^2 a_{[3,i]} + z_5^2 z_6^2 a_{[3,i]} + z_2 z_3 z_5 z_6a_{[4,i]}) 
\, , \nonumber \\   
 s_j^{(-)}= &z_1 z_4 (z_2 z_3 - z_5 z_6) (z_2 z_3 + z_5 z_6) b_{[1,j]}+ 
 z_1^2 z_2 z_6 (z_2 z_3 b_{[2,j]} + z_5 z_6 b_{[3,j]})  \nonumber \\ &  -z_3 z_4^2 z_5 (z_5 z_6 b_{[2,j]} + z_2 z_3 b_{[3,j]})  \, .
\end{align}
 Here $a_{[m,i]}$ and $b_{[n,j]}$ are generic complex constant coefficients. It can readily be checked that all fixed points miss the specialized hypersurface equation and hence, the quotient geometry is smooth.
 
The $dP_3$ base exhibits four divisor classes $h^{1,1}(B)=4$ on the covering geometries. The identification $(z_i \leftrightarrow z_{i+3})$ fixes one overall K\"{a}hler class \cite{Braun:2017feb} resulting in $h^{1,1}(dP_3/\mathbb{Z}_2)=3$ while it leaves the ambient space classes of $\mathbb{F}_0$ invariant. 
  The quotient threefold $\tilde{X}$ admits the Hodge numbers
\begin{align}
(h^{1,1},h^{2,1})_{\chi}(\tilde{X}) = (5,29)_{-48} \, .
\end{align}

\subsubsection{The effective physics}

We are now left with the task of comparing the physical theories associated to the upstairs geometry ($X$) and downstairs geometry ($\tilde{X}$) along the lines of the discussion in Section \ref{sec:quotients}. At this point, the genus one nature of the covering space geometry becomes important and it is worth a brief digression here to explain our philosophy in such cases. We will interpret the $6$-dimensional physics associated to any genus one fibered CY 3-fold via its Jacobian following standard techniques \cite{Klevers:2014bqa}. In the context of imposing discrete symmetries then we have a commutative diagram of the form

\begin{centering}
\begin{equation}
  \begin{array}{lllll}
  &X &\stackrel{\phi}{\longrightarrow}&J(X)&\\
  \Gamma &\downarrow&&\downarrow&\Gamma \\
  &\tilde{X}&\stackrel{\tilde{\phi}}{\longrightarrow}&J(\tilde{X})&
 \end{array}\label{jac_quot}
\end{equation} 
\end{centering}
where we will practically construct the ``quotient'' action on the Jacobian, $J(X)$, via imposing equivariance of the defining equations of $X$ and then mapping these across the morphism $\phi$ above to produce a restricted form of that Jacobian which will determine the form of the Jacobian, $J(\tilde{X})$ of the genus one fibered manifold $\tilde{X}$.

The analysis of $J(X)$ for the geometry chosen above associates to this genus-one fibration a $\mathbb{Z}_2 \times U(1)$ gauge group, as has been investigated in \cite{Klevers:2014bqa} (with general formulas for the spectrum computation). The full spectrum of covering and quotient theories are summarized in Table~\ref{tab:example1} and consists of several U(1) and discrete charged multiplets.
\begin{table}
\begin{center} 
\begin{tabular}{|c|c|c|}\hline
Group & \multicolumn{2}{|c|}{$\mathbb{Z}_2 \times U(1)$} \\ \hline
Multiplicity & $X$ & $\tilde{X}$ \\ \hline \hline 
$\mathbf{1}_{(0,-)}$& 60 & 30 \\ \hline
$\mathbf{1}_{(1,-)}$& 36 & 18 \\ \hline
$\mathbf{1}_{(1,+)}$& 36 & 18 \\ \hline \hline
$\mathbf{1}_{(0,+)}$ & 55 & 30 \\ \hline 
$\mathbf{V}$ & 1 & 1 \\ \hline
$\mathbf{T_{(1,0)}}$ & 3 & 2 \\ \hline
$\mathbf{T_{(2,0)}}$ & 0 & 4 $\times \mathcal{A}_1$ \\ \hline
\end{tabular}
\caption{\label{tab:example1}{\it Summary of the massless 6d spectrum of the covering and quotient threefold $X$ and $\tilde{X}$. The quotienting does not change the gauge group but does affect the tensor content of the base.}} 
\end{center}
\end{table}
We find that the overall gauge symmetry stays invariant in this case while the quotient produces four $\mathcal{A}_1$ singularities in the base with multiple fibers over them. The quotient acts freely on the multiplicity of hypermultiplets, dividing them by $1/2$ as none of them lie on fixed point loci. In this example, by using the form of the Jacobian of the genus one fibered geometry and the reasoning laid out in \eqref{jac_quot} above, the spectrum can be readily determined via the discussion in Section \ref{sec:quotients} and can be verified to be anomaly free by including the appropriately reduced $(1,0)$ tensors. 

\subsection{U(1) reducing $\mathbb{Z}_2$ quotient}\label{u1_eg}
Below we will provide our first example of a quotient which reduces the rank of the total gauge group. Geometrically the quotient action globally {\it identifies sections} generating the free part of the Mordell-Weil group with the zero-section, leading to a downstairs CY geometry that is genus-one fibered (and includes multiple fibers). The form of this identification is particularly easy to see over fixed points in the base as we will illustrate below.

\subsubsection{The geometry}
For ease of exposition we choose the same geometry as in the example before, given by the polytope \eqref{eq:polyexample2} but switch the role of the fiber and base ambient space.
\begin{table}[h!]
\begin{center}
%\begin{tabular}{cc}
$
\begin{array}{|c|c|c|}\hline
& X & \tilde{X} \\ \hline  
\chi& -96 & -48 \\ \hline
h^{1,1}& 6 & 5 \\ \hline
h^{2,1}& 54 & 29 \\ \hline 
$Group$& U(1)^3 & U(1)^2 \times \mathbb{Z}_2 \\ \hline
 V & 3 & 2 \\ \hline
 H_{\mathbf{1}_q}& 192 & 96 \\ \hline
T_{1,0} & 1 & 1 \\ \hline
T_{2,0} & 0 & 4 \times \mathcal{A}_1 \\ \hline
\end{array}$
\end{center} 
\caption{\label{tab:SpectrumExample3}{\it Summary of the F-theory matter content associated to a CY covering geometry $X$ and its quotient $\tilde{X}$ with reduced Mordell-Weil rank.}}
\end{table}
In this case then the elliptic fiber is described by an equation of the form
\begin{align}
p=&s_1^{(-)} z_5 z_3^2 z_4^2 z_2 + s_2^{(+)} z_1 z_3^2 z_4 z_2^2 + s_3^{(-)} z_5^2 z_3 z_4^2 z_6 + 
 \widehat{s}_4 z_5 z_1 z_3 z_4 z_2 z_6 \nonumber \\ &+ s_3^{(+)} z_1^2 z_3 z_2^2 z_6 + s_2^{(-)}  z_5^2 z_1 z_4 z_6^2 + 
 s_1^{(+)} z_5 z_1^2 z_6^2 z_2 \, .
\end{align}
within the ambient space $dP_3$ as defined in \eqref{eq:polyexample2}. In that space each toric ray is a $-1$ curve which results in a rational section on the generic fiber \cite{Braun:2013nqa}. To begin, the $s_i$ can be taken to be generic polynomials in the anti-canonical class of the $\mathbb{F}_0$ base. Upon fixing a zero-section, only three of the rational sections are linearly inequivalent under the Mordell-Weil group law \cite{Braun:2013nqa} for this hypersurface realization of the elliptic fiber. One such choice in the $z_i$ is given as
\begin{align}
\begin{split}
S_0: p=&z_4=0: [1,1,s_1^{(+)},0,-s_3^{(+)},1] \, , \\
S_1: p=&z_5=0: [1,1,1,s_3^{(+)},0,-s_2^{(+)}] \, , \\
S_2: p=&z_1=0: [0,-s_3^{(-)},1,1,1,s_1^{(-)}] \, , \\
S_3: p=&z_3=0: [1,s_2^{(-)} ,0,-s_1^{(+)},1,1] \, .
\end{split}
\end{align}

The effective theory associated to the upstairs geometry in this case must contain a $U(1)^3$ gauge group. Using the general formulas outlined in
 \cite{Klevers:2014bqa}, the full charged spectrum can be determined and is given in  Table~\ref{tab:SpectrumExample3}. 
 
It now remains to impose the $\mathbb{Z}_2$ symmetry on $X$ described in the previous Subsection (see e.g. \eqref{eq:qutientEx2}). As above, we can view this process as a specialization in complex structure of the $s_i$. These $s_i$ come in the general form
\begin{align}
s_i^{(\pm)}=&x_0^2 (y_1^2 a_{[1,i]} + y_0^2 a_{[2,i]} \pm y_0 y_1 a_{[3,i]}) \pm
 x_0 x_1 (y_1^2 a_{[4,i]} \pm y_0 y_1 a_{[5,i]} + y_0^2 a_{[6,i]}) \nonumber \\ &+ 
 x_1^2 (y_1^2 a_{[7,i]} + y_0^2 a_{[8,i]} \pm y_0 y_1 a_{[9,i]}) \, , \\
\widehat{s}_4=&
x_0^2 y_1^2 b_{[1]} + x_0^2 y_0^2 b_{[2]} + x_0 x_1 y_0 y_1 b_{[3]} + 
 x_1^2 y_1^2 b_{[4]} + x_1^2 y_0^2 b_{[5]}\, ,
\end{align}
with $a_{[i,j]}, b_{[i]}$ being generic complex constants.
Note that only $\widehat{s}_4$ is actually a $\mathbb{Z}_2$ invariant polynomial, in contrast to the others (denoted with superscripts) which are interchanged under the $\Gamma_{2,b}$ action $s^{(+)}_i \leftrightarrow s_i^{(-)}$. Note that the above transformation behavior under $\Gamma_{2,b}$ does not pose a problem for the associated Weierstrass model as the relevant objects, such as 
\begin{eqnarray}
f=-\frac13\left((s_1^{(-)})^2 (s_1^{(+)})^2 + (s_2^{(-)})^2 (s_2^{(+)})^2 - s_2^{(-)} s_2^{(+)} s_3^{(-)} s_3^{(+)} + (s_3^{(-)})^2 (s_3^{(+)})^2 \right. \\  \left. -
       s_1^{(-)} s_1^{(+)} (s_2^{(-)} s_2^{(+)} + s_3^{(-)} s_3^{(+)})\right)     \nonumber    -\frac12\left( 
    (s_1^{(+)} s_2^{(+)} s_3^{(-)} + s_1^{(-)} s_2^{(-)} s_3^{(+)}) \widehat{s}_4  \right.\\  \left. \nonumber+ 
   8 (s_1^{(-)} s_1^{(+)} + s_2^{(-)} s_2^{(+)} + s_3^{(-)} s_3^{(+)}) (\widehat{s}_4)^2 - (\widehat{s}_4)^4\right) \, ,
 \end{eqnarray}  
 and similarly $g$, are manifest $\Gamma_{2,b}$ invariant sections. 

\subsubsection{The Effective Physics}
For this fibration, it is clear that $S_0$ and the Section $S_2$ get interchanged upon the $\Gamma_{2}$ action \eqref{eq:qutientEx2}. Another good consistency check of this can be obtained by considering the mapping of points on the elliptic fiber over a fixed point in the base where the intersection points of the two sections are manifestly mapped into one another. 

This mapping of sections clearly has repercussions for the Shioda map and the generator of the U(1) gauge symmetry, given as
\begin{align}
\sigma(S_2)= [S_2]-[S_0]-K^{-1}_{ \mathbb{F}_0 } \, .
\end{align}
Under the identification of $S_2$ and $S_0$ this becomes trivial resulting in a loss of free Mordell-Weil rank and zero-section\footnote{Since $f$ and $g$ are $\Gamma_{2,b}$ invariant sections, it is clear that also this model becomes singular if the sections are not shifted accordingly.}.
The quotient action does not reduce the dimension of the cohomology of the $\mathbb{F}_0$ base and thus preserves the number of $(1,0)$ tensors present in the covering theory.  In addition there are again the four $\mathbb{Z}_2$ fixed points that augment the 6-dimensional SUGRA theory with four $\mathcal{A}_1$ discrete gauged subsectors.

As described in Section \ref{sec:quotients}, the quotient again acts freely on the matter multiplets, as the fiber is smooth over all fixed points, reducing their multiplicities simply by one half. The full spectrum is summarized in Table~\ref{tab:SpectrumExample3} which is manifestly consistent with anomaly cancellation.

\subsection{Non-Abelian group reducing quotients}
\label{sec:nonAbReducing}
In this section we present quotients which involve the identification of sections in an elliptically fibered covering space geometry and as described in Section \ref{sec:quotients}, torsional sections of the fibration. In each case the quotient action will identify fibral divisors in such a way that there is no residual continuous gauge group associated to the downstairs genus-one fibered geometries. However, these kinds of quotients are potentially interesting as they allow to systematically construction models with (possibly high order) discrete gauge groups of the same order as the torsional Mordell-Weil group. We will refer to these again in Section~\ref{sec:bounds} in order to comment on potential bounds to the order of discrete symmetries in six dimensions.

\subsubsection*{An $SU(2)/\mathbb{Z}_2$ quotient}
The simplest possible example, that of an $SU(2)/\mathbb{Z}_2$ gauge group, can be obtained from a Weierstrass model with a $\mathbb{Z}_2$ torsion point \cite{Aspinwall:1998xj}, given as 
\begin{align}
\label{eq:WSFZ2}
y^2 =& x(x^2 + a_2 x + a_4) \, , \qquad a_4 \in \mathcal{O}(K_b^{-4})\, , \quad a_2 \in \mathcal{O}(K_b^{-2})\, . \\
f=& a_4 - \frac13 a_2^2 \, , \quad g= \frac{1}{27} a_2 (2 a_2^2 -9 a_4) \\
\Delta=& a_4^2 (4 a_4 - a_2^2) \, 
\end{align}
The $\mathbb{Z}_2$ torsion point in this representation is located at $y=x=0$, which is visible due to its  $y \rightarrow -y$ symmetry. As argued before, this model requires {\it by construction} at  least an SU(2) gauge factor over the $a_4=0$ locus. As required, the collision of $a_4=0$ with the $I_1$ locus produces an order $V(f,g,\Delta)=(1,2,3)$ singularity which leads to no additional matter. Hence the charged hypermultiplet sector of the theory contains adjoint matter only and is counted by the genus of the $a_4=0$ curve. \\

For simplicity, we choose to consider the tuned Weierstrass model above over the base $\mathbb{F}_0$ (chosen since it's suitable for a $\mathbb{Z}_2$ quotient). This complete Calabi-Yau threefolds has a simple toric description in terms of the following polytope 
\begin{align}
\begin{array}{|cccc|cccc|}\hline
x_0 & x_1 & y_0 & y_1 & u & v & w &  e_1 \\ \hline
1&-1&0&0& 0   &0&0&0\\ 
0&0&1&-1&0    &0&0&0\\
-2&-2&-2&-2&-2&0&2&1\\
-1&-1&-1&-1&-1&1&-1&0\\ \hline
\end{array}_{(h^{1,1}=4,h^{2,1}=148)}^{(\chi=-288)} \, ,
\end{align}
where the superscript and subscripts above are the Euler number and Hodge numbers respectively. The Calabi-Yau is given as the anti-canonical hypersurface with defining equation
\begin{align}
p= b_1 u^4 + b_2 u^2 w^2 e_1 +   w^2  e_ 1^2 + b_6 u w v e_ 1 + 
 e_ 1 v^2 \, ,
\end{align}
that admits the Weierstrass form \eqref{eq:WSFZ2} upon the identification
\begin{align}
\begin{split}
f=&b_1 - 1/48 (-4 b_2 + b_6^2)^2 \, , \\
g=&1/864 (4 b_2 - b_6^2) (72 b_1 - (-4 b_2 + b_6^2)^2)\, ,\\
\Delta=&1/16 b_1^2 (64 b_1 - (-4 b_2 + b_6^2)^2)\, ,\\
\end{split}
\end{align}
which can be obtained from the generic model upon shifting
\begin{align}
a_2 \rightarrow  (- b_2 + 1/4 b_6^2)\, , \qquad  a_4 \rightarrow b_1 \, .
\end{align}
 We fix a triangulation of the polytope with SRI
\begin{align}
SRI: \{ x_0 x_1, y_0 y_1, u e_1, v w \} \, .
\end{align}
We choose $u=0$ as the zero section of the fiber, and $v$ (or equivalently $w$) as the $\mathbb{Z}_2$ torsion point which intersect the $SU(2)$ resolution divisor $D_{e_1}=0$ (see \cite{Klevers:2014bqa} for a detailed study of this fiber type). The particle spectrum associated to the $6$-dimensional $SU(2)/\mathbb{Z}_2$ upstairs theory is given in Table~\ref{tab:SU2Spectrum}.
\begin{figure}[t!]
\begin{center}
\includegraphics[scale=1.5]{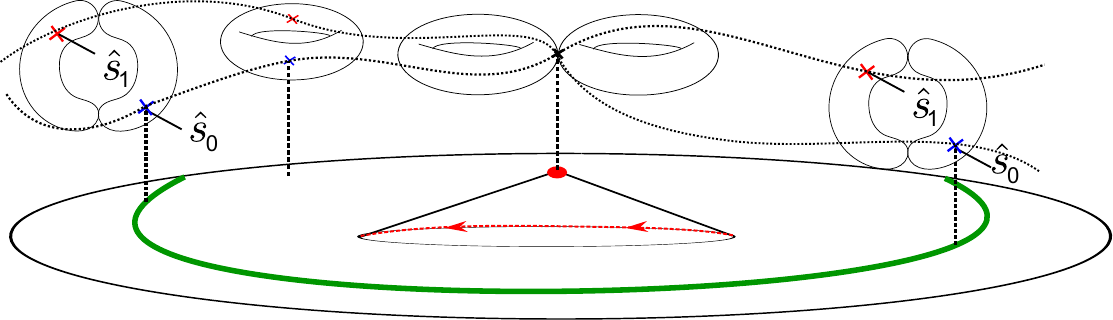}
\caption{\label{fig:SU2Torsion}{\it Depiction of the $SU(2)/\mathbb{Z}_2$ fibration after taking the quotient. The two sections $\hat{s}_0$ and $\hat{s}_1$ intersect the two SU(2) irreducible curves. Their identification over the multiple fiber locus breaks the SU(2) at codimension one.} }
\end{center}
\end{figure}

This upstairs geometry admits a quotient by a freely acting $\mathbb{Z}_2$ symmetry. The $\mathbb{Z}_2$ action ($\Gamma_2$) on the threefold in terms of fiber coordinate $\omega$ and $\mathbb{P}^1_{x} \times \mathbb{P}^1_y$ coordinates can be given as
\begin{align}
\label{eq:Z2action}
\Gamma_2: (x_0, x_1, y_0, y_1; \omega) \rightarrow (x_0, -x_1, y_0, -y_1; \omega+\hat{s}_1)\,,
\end{align} 

The smooth quotient threefold, $\tilde{X}$ admits the following Hodge numbers \cite{Donagi:1999ez,Donagi:2000zf,Braun:2017feb}
\begin{align}
(h^{1,1},h^{2,1})_\chi (\tilde{X}) = (3,75)_{-144} \, ,
\end{align}
 which is now a two-section genus fibered geometry with four $\mathcal{A}_1$ fixed points in the base and multiple fibers over them. A picture of the geometry is given in Figure~\ref{fig:SU2Torsion}. Note that the covering geometry only admitted matter in the form of adjoint charged hypermultiplets over the genus 49 curve $a_4=0$. The smooth quotient reduces this number to 
 \begin{align}
 \mathbf{3} \times (1+48) \rightarrow 24 \times ( \mathbf{1}_1 \oplus \mathbf{1}_{-1} ) \, ,
 \end{align}
 that is $48$ discrete charged hypers. The addition of the four $\mathcal{A}_1$ discrete charged SCFT points completes the full spectrum, as summarized in Table~\ref{tab:SU2Spectrum},(and is consistent with anomaly cancellation, as expected). 
 \begin{table}[h!]
\begin{center} $
\begin{array}{|c|c|c|}\hline
& X & \tilde{X} \\ \hline  
\chi& -288 & -144 \\ \hline
h^{1,1}& 4 & 3 \\ \hline
h^{2,1}& 148 & 75 \\ \hline 
$group$ & SU(2)/\mathbb{Z}_2& \mathbb{Z}_2 \\ \hline
V & 3 & 0 \\ \hline
H& \begin{array}{l}\mathbf{1}_0 \times 149\\
\mathbf{3}\times 49
  \end{array} 
 & \begin{array}{l}\mathbf{1}_0 \times 76
 \\
 \mathbf{1}_1 \times 48  
  \end{array}\\  \hline
T_{1,0} & 1 & 1 \\ \hline
T_{2,0} & 0 & 4 \times \mathcal{A}_1 \\ \hline
\end{array}$ 
\caption{\label{tab:SU2Spectrum}{\it Summary of the 6-dimensional F-theory spectrum on the threefold, $X$, and its quotient, $\widetilde{X}$.}}
\end{center} 
 \end{table}

\subsubsection*{Example of an $(SU(3)\times SU(3))/\mathbb{Z}_3$ quotient}

Moving on from the simple example of the previous subsection, we next turn to an example with two non-Abelian gauge group factors and a $\mathbb{Z}_3$ torsion point.  The tuned Weierstrass form of the most general $\mathbb{Z}_3$ torsion model is given as  \cite{Aspinwall:1998xj}
\begin{align}
\begin{split}
y^2+& a_1 xy+a_3 y =  x^3 \, , \quad a_1 \in \mathcal{O}(K_b^{-1})\, , a_3 \in \mathcal{O}(K_b^{-3}) \\
f=& \frac12 a_1 a_2 - \frac{1}{48} a_1^4 \, , \quad
g=  \frac{1}{4} a_3^2    +\frac{1}{864}a_1^6 - \frac{1}{24} a_1^3 a_3 \, , \\ 
\Delta=& \frac{1}{16} a_3^3 (27 a_3 - a_1^3) \, .
\end{split}
\end{align}
The $a_3=0$ divisor signals the presence of at least one $SU(3)/\mathbb{Z}_3$ gauge factor if it does not become reducible. In the following we consider a splitting of the form $a_3 \rightarrow b_2 b_1$ which yields a $(SU(3)\times SU(3))/\mathbb{Z}_3$ gauge group. 

This tuned elliptic fiber can be readily found within a smooth toric hypersurface which describes an elliptic fibration over a $\mathbb{P}^2$ base. This Calabi-Yau hypersurface is given by the following polytope 
\begin{align}
\label{eq:polyexample2a}
\begin{array}{|ccc|ccccccc|}\hline
x_0 & x_1 & x_2& u & e_1 & e_2 &  v & w  & e_3 & e_4 \\ \hline
1    & 0     &  -1 & 0    & 0     &    0 &    0 &   0 & 0   &0 \\
0    &   1   &  -1 & 0   &  0    &    0 &   0  &  0   & 0 & 0 \\
-1   &  -1  &  -1 &  -1  & 0   &    1  &  2  &  -1  & 0 & 1\\
0    & 0    &   0  &   1 &  1   &    1 &  1  & -2    &-1 &0\\ \hline 
\end{array}_{(h^{1,1}=6,h^{2,1}=60)}^{(\chi=-108)} \, .
\end{align}
where once again the superscript and subscripts denote Euler and Hodge numbers. The $\mathbb{P}^2$ base is given by the projection onto the first two columns. There exist two triangulations of the ambient toric variety. Here we consider one that leads to the Stanley-Reisner ideal
\begin{align}
SRI: \{& u e_2, u v, u w, u e_3, u e_4, e_1 w, e_2 w, v w, w e_4, e_1 v, e_1 e_3, e_1 e_4,  e_2 e_3, \nonumber \\ & e_2 e_4, v e_3, x_0 x_1 x_2 e_1, x_0 x_1 x_2 e_2,   x_0 x_1 x_2 v, x_0 x_1 x_2 e_3, x_0 x_1 x_2 e_4 \} \,.
\end{align} 
   The three sections $u,v,w$ admit a $\mathbb{Z}_3$ torsional relation \cite{Braun:2013nqa} and the $e_i$ are associated to the four resolution divisors of the two SU(3) gauge groups. The hypersurface equation is given as
\begin{align}
p=b_1 e_1^2   e_2 u^3 + a_1 e_1  e_2 v w u e_3 e_4 + b_2  w^3 e_3^2 e_4 +  
 e_1 e_2^2 v^3 e_3 e_4^2\, ,
\end{align}
where $\{e_1 , e_2\}$ and $\{ e_3,e_4\}$ correspond to the resolution divisors over $b_2=0$ and $b_1=0$ that are degree six and degree three polynomials in the base $\mathbb{P}^2$. Choosing $v=0$ as the zero section, the other two sections $w=0$, $u=0$ combine with the zero section to admit the $\mathbb{Z}_3$ torsion relation. From the toric diagram it becomes directly visible that each torsion section intersects one of the three irreducible fiber components each.
 \begin{table}[h!]
\begin{center} $
\begin{array}{|c|c|c|}\hline
& X & \tilde{X} \\ \hline  
\chi& -108 & -36 \\ \hline
h^{1,1}& 6 & 2 \\ \hline
h^{2,1}& 60 & 20 \\ \hline 
$Group$ & SU(3)^2/\mathbb{Z}_3& \mathbb{Z}_3 \\ \hline
V & 16 & 0 \\ \hline
H& \begin{array}{l}\mathbf{1}_0 \times 61 \\
(\mathbf{3},\overline{\mathbf{3}})\times 18\\
(\mathbf{8},\mathbf{1})+(\mathbf{1},\mathbf{8})\times 10 
  \end{array} 
 & \begin{array}{l}\mathbf{1}_0 \times 21
 \\
 \mathbf{1}_1 \times 9\cdot 6 \\
 \mathbf{1}_{1}\times  3 \cdot 6 
  \end{array}\\  \hline
T_{(1,0)} & 0 & 0 \\ \hline
T_{(2,0)} & 0 & 3 \times \mathcal{A}_2 \\ \hline
\end{array}$ 
\caption{\label{tab:SU3Spectrum}{\it Summary of the spectrum of the $SU(3)^2/\mathbb{Z}_3$ theory obtained from geometry $X$ and its quotient theory on $\widetilde{X}$.}}
\end{center} 
 \end{table}

To realize a freely-acting $\mathbb{Z}_3$ quotient, the standard toric $\mathbb{Z}_3$ action on the base $\mathbb{P}^2 : x_j \sim e^{2\pi i /3 j}x_j$ can be combined with an order three shift of the sections analogous to \eqref{eq:Z2action}.

Under this quotient action the sections are identified and the resolution divisors corresponding to $SU(3)$ roots will be identified with the affine one. As a result, the quotient is associated to a downstairs theory with fully broken continuous gauge group and a residual $\mathbb{Z}_3$ symmetry. This expectation can be verified by considering the upstairs and downstairs matter spectrum which is listed in Table~\ref{tab:SU3Spectrum} which is consistent with all anomalies.

\subsection{Combining effects in a $\mathbb{Z}_6$ quotient}
With the observations of the previous two subsections in hand, we can attempt to engineer an example with a higher order discrete symmetry group. While examples of geometries with $\mathbb{Z}_3$ and $\mathbb{Z}_4$ have appeared before \cite{Klevers:2014bqa,Cvetic:2015moa,Braun:2014qka,Oehlmann:2016wsb}, higher order discrete symmetries have proven more elusive. In this section we discuss an example of a non-prime quotient that reduces the numbers of tensors and also the number of Abelian gauge group factors. The geometry is realized as a complete intersection CY threefold \cite{Candelas:1987kf, Anderson:2017aux,Anderson:2018kwv} given by the configuration matrix.
\begin{align}
M = \left[ \begin{array}{c|cccc}   
\mathbb{P}^1_x & 1 & 0 & 0 & 1\\     
\mathbb{P}^1_y & 1 & 0 & 0 &1  \\     
\mathbb{P}^1_z & 1 & 0 & 0 & 1\\
      \mathbb{P}^2_u & 0 & 1 & 1 & 1\\
      \mathbb{P}^2_w & 0 & 1 & 1 & 1\\
                              \end{array} \right]_{(h^{1,1}=8,h^{2,1}=44)}^{(\chi=-72)}
\end{align}
%with configuration number $\# 7206$ $
where here the first column denotes an ambient space which is a product of projective space factors and each column denotes the multi-degree of an equation in this space defining the complete intersection. As in other examples, super-/sub-scripts denote topological data.

Within this description, the base of the fibration can be identified as $dP_3$ realized as a hypersurface
\begin{align}
dP_3 = \left[ \begin{array}{c|c} 
  \mathbb{P}^1_x & 1 \\
  \mathbb{P}^1_y & 1 \\
\mathbb{P}^1_z & 1
\end{array} \right]~.
\end{align}
For this CY threefold, the genus one fiber is given as
\begin{align}
\text{genus-one fiber} \sim \left[ \begin{array}{c|ccc}
   \mathbb{P}^2_u & 1 & 1 & 1\\
      \mathbb{P}^2_w & 1 & 1 & 1\\
      \end{array} \right] \, ,
\end{align} 
that does not admit a section, but three-sections only. 

Finally it should be noted that the Jacobian of this genus-one fibration can be readily constructed and leads to a rank three Mordell-Weil group in that elliptically fibered geometry. Hence, in this case the upstairs CY geometry is associated to a $U(1)^3$ gauge group. 

This threefold has a known, freely acting $\mathbb{Z}_6$ symmetry \cite{Braun:2010vc}, acting as 
\begin{align}
\begin{split}
\Gamma_{6,f}:& (x_0\rightarrow y_0,   x_1 \rightarrow  -y_1, y_0 \rightarrow z_0, y_1 \rightarrow-z_1, z_0 \rightarrow x_0, z_1 \rightarrow-x_1) \, ,\\
\Gamma_{6,b}: &( u_0 \rightarrow w_0,u_1 \rightarrow w_1 \gamma_3^2 ,  u_2 \rightarrow w_2 \gamma_3,  w_0   \rightarrow u_0,  w_1 \rightarrow u_1,  w_2 \rightarrow u_2   ) \, ,
\end{split}
\end{align}
with $\gamma_3$ a third root of unity, including the non-trivial action on the hypersurface equations
\begin{align}
\Gamma_6: &(b_1,f_1, f_2, f_3) \rightarrow (b_1,f_1 ,\gamma_3^2 f_2, -\gamma_3 f_3) \, ,
\end{align} 
reducing the Hodge numbers of the quotient \cite{Constantin:2016xlj}, $\widetilde{X}$ to
\begin{align}
(h^{2,1}(\widetilde{X}),h^{1,1}(\widetilde{X}))=(2,8) \, .
\end{align}

Within the quotient CY threefold, the fixed points in the base can be found by considering a $\Gamma_{6,b}$-invariant equation for the base
\begin{align}
\hat{b} = a_{000} x_0 y_0 z_0 + a_{110} (x_1 y_1 z_0 + x_1 y_0 z_1 + x_0 y_1 z_1) \, .
\end{align}
Within this description, we find a set of $\Gamma_{6,b}^i$ fixed points of orders two, three and six. Note that several of them get identified upon residual $\mathbb{Z}_6$ elements. These fixed points are summarized in Table~\ref{tab:Z6fixedpoints}.
\begin{table}[t!]
\begin{center}
\begin{tabular}{|c|l|c|}\hline
\text{Order} & $(x_0,x_1;y_0,y_1;z_0,z_1)-(\mathbb{P}^1)^3-\text{Coordinate}$& \text{multiplicity} \\ \hline
$\mathbb{Z}_6$&  $(0,1;0,1;0,1)$ & 1\\ 
$\mathbb{Z}_3$ & $(\sqrt{3 a_{110}}, \pm i\sqrt{a_{000}};\sqrt{3 a_{110}},\pm i\sqrt{a_{000}};\sqrt{3 a_{110}},\pm i\sqrt{a_{000}})/\sim_{\mathbb{Z}_2}$& 1\\
$\mathbb{Z}_2$ &$ \{ (0,1;1,0;1,0) - (1,0;0,1;1,0) - (1,0;1,0;0,1) \}/\sim_{\mathbb{Z}_3}$ & 1  \\ \hline
\end{tabular}
\caption{\label{tab:Z6fixedpoints}{\it Summary of $\mathbb{Z}_6$ fixed points in the base. The residual identification identifies several solutions, leaving only one fixed point of each order.      } }
\end{center}
\end{table} 

In terms of the physical theory, once again the downstairs geometry is associated to a discrete gauge group only. The symmetry action identifies sections and base divisors such that the number of vectors and tensors in the downstairs theory is fully reduced. The action on the base produces three fixed points of orders two, three and six, respectively in the base with multiple fibers of the same orders over them. We thus deduce that the geometry presented above is associated to a $\mathbb{Z}_6$ discrete gauge symmetry with 24 discrete charged singlets. The full spectrum of covering and quotient theory is summarized in Table~\ref{tab:SpectrumZ6}.
\begin{table}[ht!]
\begin{center}
\begin{tabular} {|l|c|c|}\cline{2-3}
\multicolumn{1}{c|}{}& Covering & Quotient \\ \cline{1-3}
Gauge Group& $U(1)^3 \times \mathbb{Z}_3$ &$ \mathbb{Z}_6$ \\ \cline{1-3}
 & \multicolumn{2}{c|}{\text{multiplicity}} \\ \hline
$H_{\text{neut}} $& 45& 9 \\
$H_{\text{charged}} $& 144 & 24 \\
$V $& 3& 0 \\
$T_{(1,0)} $& 3 &0 \\
$T_{(2,0)}$ & 0 & $\mathcal{A}_1 \oplus \mathcal{A}_2 \oplus \mathcal{A}_5$\\ \hline 
\end{tabular}
\caption{\label{tab:SpectrumZ6}{\it Summary of the spectrum of a $U(1)^3$ CICY model and its $\mathbb{Z}_6$ quotient.}}
\end{center}
\end{table}
Note that, subject to the {\it Working Assumption} mentioned in Section \ref{sec:intro}, to our knowledge, this is the first time that an order six discrete symmetry has been constructed in six dimensional F-theory compactifications.

\section{Quotients of the Schoen manifold}
\label{sec:bounds}
The discussion in the previous sections has involved the general properties that can arise in quotients of elliptically fibered Calabi-Yau threefolds, however it is hampered by the fact that no complete classification of such symmetries is yet known (see \cite{Braun:2010vc,Braun:2017juz} for systematic efforts with some data sets of manifolds). In this Section we consider one manifold for which all possible (fibration preserving) discrete symmetries have been classified \cite{Bouchard:2007mf} -- the so-called ``Schoen" or ``split bi-cubic" threefold with Hodge numbers $(h^{11},h^{21})=(19,19)$. 

In particular, as in Section~\ref{sec:nonAbReducing}, in this section we will systematically consider fibrations with higher order Mordell-Weil torsion. Realizing these fibrations within the Schoen manifold will allow us to use it as the covering space of quotient CY threefolds associated to theories with higher order discrete symmetries. As with freely acting discrete symmetries, a full classification of the possible MW torsion groups of elliptic threefolds would be desirable but is unfortunately an open problem \cite{Hajouji19}. To begin, it is worth noting that Mazur \cite{Mazur78} has classified torsion groups for a single elliptic curve over  $\mathbb{Q}$, and found the following:
\begin{align}
\mathbb{Z}_k \, , n = 1,\ldots 10,12 \, ,\quad  \mathbb{Z}_2 \oplus \mathbb{Z}_m, \, ~m = 2,4,6,8 \, .
\end{align}
Moreover, for elliptic K3 surfaces, a MW torsion classification also exists \cite{Shimada00} with orders $n=2\ldots 8$ as well as $m=2,4,6$ but also $\mathbb{Z}_3\oplus \mathbb{Z}_3$ and $\mathbb{Z}_4 \oplus \mathbb{Z}_4$ and hence are not simply included in the set of Mazurs classification. However, for CY three and fourfolds, such a classification is unknown. 

For CY threefolds, a classification of MW torsion could be used to systematically construct higher order discrete symmetries by taking a sufficient quotient of the theory (as described in previous sections). The state of the art in constructing explicit Weierstrass models with additional torsional points of various orders was  performed by Aspinwall and Morrison \cite{Aspinwall:1998xj}, with models ranging from
\begin{align}
\mathbb{Z}_n\, , n=2, \ldots 6 \, , \qquad \mathbb{Z}_2 \oplus \mathbb{Z}_{2m} \, , m=2,4 \, , \qquad \mathbb{Z}_3 \oplus \mathbb{Z}_3 \, .
\end{align} 
In the following we will take the models of Aspinwall and Morrison and consider CY quotients acting via rotation of sections within those torsion groups in the fiber and with a non-trivial action in the base of the fibration.

In all known examples the discrete action in the fiber and the base are the same group. Thus it is clear that the case symmetry action also constrains the possible symmetries appearing in the fiber (i.e. torsion groups) for this class of models. For a Fano base for example, the order of the quotient is already restricted purely from the consideration of the reducible gravitational anomaly
\begin{align}
9-T_{(1,0)}= (K_b^{-1})^2 \, ,
\end{align}
where $9>(K_b^{-1})^2 > 0$ and in addition, both sides must be divisible by the order $n$ and thus at most an order nine quotient is possible. 

In the case of the Schoen threefold, $\pi: X  \to dP_9$ and this base surface seems naively, to allow for infinite order quotients. However this is not the case and all freely acting discrete symmetries were classified in  \cite{Zhang00,Bouchard:2007mf} and in fact, do not exceed the orders above. In this section we consider these higher order torsion models and discuss their F-theory physics as wells as their quotients. 

The Schoen manifold is well-known to be an exceptional point in the landscape of Calabi-Yau manifolds. As a fiber product of two rational elliptic surfaces it has a range of remarkable features, including a vast number of freely acting discrete symmetries and in fact, an infinite number of inequivalent genus-one fibrations \cite{oguiso,Aspinwall:1996mw,Anderson:2017aux}. For generic points in its complex structure moduli space, the Schoen manifold has a non-trivial, rank $8$ Mordell-Weil group, the highest rank explicitly known for a Calabi-Yau threefold \cite{Morrison:2016lix}.

In the following Subsections, we begin by illustrating a $\mathbb{Z}_5$ quotient of the Schoen manifold in some detail and provide a brief summary of results for other higher order quotients in Subsection~\ref{sec:SchoenQuotients} and Appendix~\ref{app:WSFtorsion}. These explicit examples illustrate some features which lead us to comment on possible bounds on discrete symmetries in Subsection~\ref{sec:Comments}.

\subsection{F-theory on a $\mathbb{Z}_5$ torsion model and its quotient} 
\label{sec:Z5Quotient}
As written in \cite{Aspinwall:1998xj}, a $\mathbb{Z}_5$ torsion model can be obtained by the following tuned Weierstrass form:
\begin{align} 
\begin{split}
 y^2+a_1 xy &+(a_1 - b_1) b_1^2 y = x^3+(a_1 - b_1)b_1 x^2\\
f=&\frac{1}{6} a_1 b_1^3  -\frac{1}{48} a_1^4 +  \frac13 a_2^2b_1^2 -   \frac13 b_1^4-\frac16 a_1^3 b_1    \\
g=&    +\frac{1}{864} ( a_1^2 - 2 a_1 b_1+2b_1^2)(a_1^4 +14 a_1^3 b_1 +26 a_1^2 b_1^2-11 a_1 b_1^3 + 76 b_1^4       ) \\
\Delta=& \frac{1}{16} (a_1^2+9a_1 b_1 - 11 b_1^2) b_1^5 ( a_1-b_1)^5 \, , 
\end{split}
\end{align} 
This gives rise to a $SU(5) \times SU(5)/\mathbb{Z}_5$ gauge group if the sections $a_1$ and $b_1$ are generic polynomials. 

Beginning with the upstairs geometry/physics, it should be observed that since the divisors supporting both $SU(5)$ factors are both in the class of the anti-canonical class of the base, they are generically curves of genus-one (and hence, will contribute one adjoint hypermultiplet each to the massless spectrum). Due to the $\mathbb{Z}_5$ quotient in the gauge group, there is no bi-fundamental matter among the two $SU(5)$ groups as one might expect from a simple adjoint breaking of $E_8$ but instead non-minimal vanishing $(V(f,g,\Delta) \sim (4,6,12)$ leads to superconformal matter points with multiplicity $n_{scp}=(K_b^{-1})^2$ (and at best non-flat resolutions over these points in the CY threefold). Since the resolution of each non-flat $(4,6,12)$ point contributes exactly one K\"{a}hler deformation \cite{Buchmuller:2017wpe,Dierigl:2018nlv} we find for a (weak) Fano base 
\begin{align}
T+n_{scp} = 9 \, , \qquad h^{(1,1)}(X)=19 \, .
 \end{align} 
With this observation and noting that each $(4,6,12)$ point contributes 29 hypermultiplets to the gravitational anomaly, one can deduce that
 \begin{align}
 H_n + H_c - V + 29 (T+n_{scp}) -273=0 \, ,
 \end{align}
Hence, any CY elliptic fibration with a weak Fano base and this fiber type must yield $19$ complex structure moduli. This is an interesting hint that the Schoen manifold (or its cousins) is a good starting point to consider such fiber types. Moreover, by taking the Schoen as our chosen elliptic fibration, the fibration over the base $dP_9$ base is flat and $(K_b^{-1})^2=0$. Hence superconformal points are avoided.

Another motivation for considering this CY manifold is that all freely acting discrete symmetries arising on it have been classified \cite{Zhang00,Bouchard:2007mf}. In the case of a $\mathbb{Z}_5$ quotient, it exists if both rational ellipic surfaces in the fiber product admit the same $\mathbb{Z}_5$ torsion automorphism given above in their fibers.
\footnote{In the classification of Schoen quotients, smoothness obtained by choosing a symmetry action on the fiber product and in particular the shared $\mathbb{P}^1$ base in such a way that singularities of one $dP_9/\mathbb{Z}_5$ miss those of the other \cite{Bouchard:2007mf}. }.

Taking a $\mathbb{Z}_5$ quotient (compatible with the torsion action) results in a manifold with reduced Hodge numbers as summarized in \eqref{eq:Z5Summary}. The $\mathbb{Z}_5$ quotient essentially identifies all $SU(5)$ resolution divisors in the fibers and analogously eight tensors in the base. Thus only a $\mathbb{Z}_5$ discrete gauge symmetry remains in the downstairs theory and two order 5 multiple fibers that restrict to two $\mathcal{A}_4$ singularities in the base.    
\begin{align}
\label{eq:Z5Summary}
\begin{array}{|c|c|c|}\cline{2-3}
\multicolumn{1}{c|}{}& $Covering Theory$ & $Quotient$ \\ \hline
G: &  SU(5)^2/\mathbb{Z}_5& \mathbb{Z}_5 \\ \hline
(h^{1,1},h^{2,1})&(19,19)& (3,3) \\ 
H_c: &  (\mathbf{24},\mathbf{1})\oplus(\mathbf{1},\mathbf{24})    & 0 \\
T_{(1,0)}: & 9 & 1 \\  
T_{(2,0)}: &0 & \mathcal{A}_4 \oplus \mathcal{A}_4  \\ \hline
\end{array}
\end{align}
In this theory there are no ordinary charged states, but the two superconformal matter points do contribute to the tensors as shown in \eqref{eq:Z5Summary} and is consistent with anomaly cancellation. 
    
 \subsection{More Schoen manifolds and their quotients} 
 \label{sec:SchoenQuotients} 
The Schoen is an intriguing playground to construct models of higher order torsion. We give a summary of the minimal gauge group over a Fano base and matter content of these models in Table~\ref{tab:SchoenSummary}. As it turns out, all of these models admit a rank 8 gauge group localized over genus one curves in the base, that hosts exactly one adjoint representation.\begin{table}[h!]
\begin{center} 
\begin{tabular}{cc}
\begin{tabular}{|cl|   }\hline
\multicolumn{2}{|c|}{Covering Theories}  \\ \hline
MW$_{\text{tor}}$& Gauge Group   \\ \hline
$\mathbb{Z}_5$ &  $ SU(5)^2/\mathbb{Z}_5$ \\
$\mathbb{Z}_6$ &$ (SU(2) \times SU(3) \times SU(6))/\mathbb{Z}_6 $ \\
$\mathbb{Z}_2 \times \mathbb{Z}_4$ &$ (SU(2)^2 \times SU(4)^2)/\mathbb{Z}_2 \times \mathbb{Z}_4 $ \\
$\mathbb{Z}_3\times \mathbb{Z}_3$ &$ SU(3)^4/\mathbb{Z}_3 \times \mathbb{Z}_3 $ \\ \hline
Content:& $\begin{array}{rc} T_{(1,0)}:&9 \\ H_{ch}:& 1\cdot \text{Adj}(G) \\  (h^{1,1},h^{2,1}):& (19,19) \end{array}$ \\ \hline
\end{tabular}
&
\begin{tabular}{|c|   }\hline
\multicolumn{1}{|c|}{Quotient Theories}  \\ \hline
$T_{(2,0)}$ Content\\ \hline 
$2 \times     \mathcal{A}_4$ \\
$ \mathcal{A}_1 \oplus \mathcal{A}_2 \oplus \mathcal{A}_5 $\\
$ 2\cdot (\mathcal{A}_1 \oplus \mathcal{A}_3)$ \\
$4 \cdot \mathcal{A}_2 $\\ \hline
$T_{(1,0)}$: 1 \\
$H_{ch}$: 0 \\
$(3,3)$ \\ \hline

\end{tabular}

\end{tabular}
 \caption{\label{tab:SchoenSummary}{\it Summary of the minimal gauge group for various higher order torsion models over $dP_9$ bases described as a Schoen manifold. In the quotient theory the gauge symmetry is fully broken to a discrete one with only superconformal matter charged under it.}}
 \end{center} 
 \end{table}  

The fact that the Schoen manifold can be viewed as a hypersurface inside $dP_9 \times dP_9$ \cite{Anderson:2017aux} allows for a simple symmetry between the fiber and base of the geometry. In the following discussion we have engineered symmetry actions in the elliptic fibers using tuned torsional Weierstrass models. In fact, free quotients exist if we choose the $dP_9$ base to admit the same torsion structure as the fibers \cite{Bouchard:2007mf}. Under quotienting by this symmetry all eight resolution divisors in the F-theory elliptic fiber are identified as well as the eight tensor multiplets in the base. Therefore the gauge symmetry is completely broken to a discrete gauge group and the identification of the tensors in the base results in discrete charged superconformal matter. \\

There are in general many quotients of the Schoen manifold possible \cite{Bouchard:2007mf} but the general construction follows simply by picking two rational surfaces with the same automorphisms and take a fiber product to ensure smoothness of the quotient. Hence, from the perspective of the covering theory, the fiber as well as the base $dP_9$ admit the same torsion structure with resolved $G=ADE$ fibers. Since the quotient collapses all  resolution divisors of the fiber, the same happens analogously to the tensors of the $dP_9$ base which are then the singular ADE points. Hence we observe that basically the ADE (resolved) structure in fiber and base is found as codimension two singularities in the base upon the quotient.
However note that in fact we had a non-simply connected total gauge group in the fiber of type $G_{\text{total}}=G/\text{Center}(G)$ of the covering theory due to the non-trivial Mordell-Weil torsion group. Hence it is tempting to speculate  whether there exists a global structure of the superconformal matter system in the base of type $G/\text{Center}(G)$ type.

\subsection{Quotients with residual gauge groups}
\label{sec:foldings}

The previous sections focused on examples where the non-simply connected gauge factor was fully removed in the quotient process. This however does not need to be the case as exemplified in the following. Qualitatively, this effect is very similar to the Dynkin diagram folding along an outer autmorphism that produces a non-simply connected gauge group. The main difference to this construction however is that there is no section in the downstairs genus one geometry and hence the induced monodromy will always affect the affine node as well. Therefore, the folding acts always on the full affine Dynkin diagram, resulting in a twisted affine algebra \cite{Bouchard:2007mf,Braun:2014oya}. Note that such Dynkin diagrams can appear in genus-one fibrations more generally and do not require the existence of a quotient construction to be realized \cite{Anderson19}.  \\ 

\begin{figure}[t!]
\begin{center}
\includegraphics[scale=0.4]{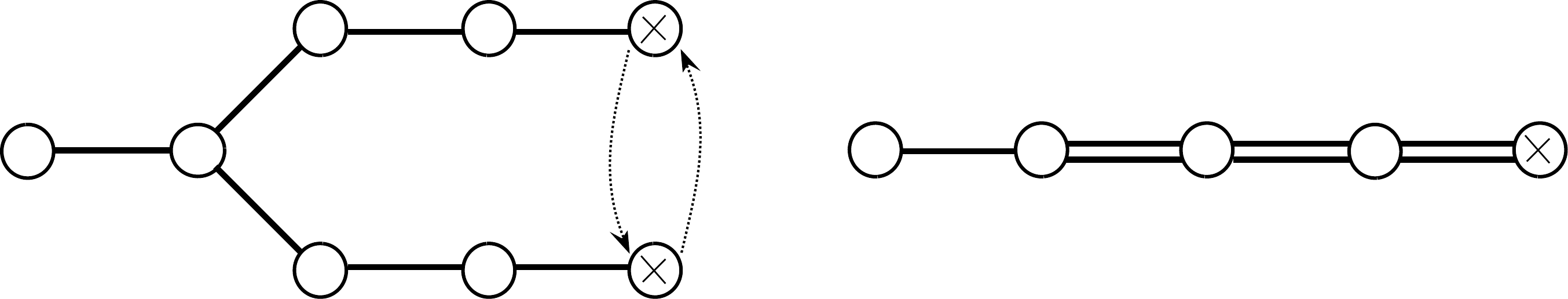}
\caption{{\it  \label{fig:F4Folding} The folding of an affine $E_7/\mathbb{Z}_2$ Dynkin diagram into that of $E^{(2)}_6$ induced by the order two monodromy of the two sections.} }
\end{center}
\end{figure}

As a starting point we start with a $\mathbb{Z}_2$ torsion model, as given in \eqref{eq:WSFZ2} and perform the additional tuning 
\begin{align}
b_4 \rightarrow b_1 c_1^3 \, ,\quad a_2 \rightarrow c_1^2\, , \qquad  [c_1], [b_1] \in K_b^{-1}\, ,
\end{align}
to obtain an $(E_7 \times SU(2))/\mathbb{Z}_2$ gauge group that are located over $c_1=0$ and $b_1=0$. In the resolution, the $\mathbb{Z}_2$ torsion section has to intersect the only multiplicity one root, as highlighted in Figure~\ref{fig:F4Folding} of the Dynkin diagrams to enforce the $\mathbb{Z}_2$ torsion factor. The smooth and flat realization of that model exists as a Schoen elliptic fibration over a $dP_9$ base. The curves $a_1=0$ and $b_1=0$ are genus one curves of self-intersection zero and hence host one adjoint, as required by anomaly cancellation of the covering theory. 
This threefold admits a free $\mathbb{Z}_2$ automorphism \cite{Bouchard:2007mf} that removes four fibral divisors, and four tensors of the base as well, adding four $\mathcal{A}_1$ superconformal theories in the base. As in the examples before the $SU(2)/\mathbb{Z}_2$ factor is broken completely. The $\mathbb{Z}_2$ acts on the $E_7$ affine Dynkin diagram by a $\mathbb{Z}_2$ folding into that of an $E_6^{(2)}$ as shown in Figure~\ref{fig:F4Folding}.  
Hodge numbers and the F-theory interpretation of the resulting gauge theories of covering and quotient theory is summarized in Table~\ref{tab:FoldingQuotient}. The gauge theory interpretation of the $E_6^{(2)}$ Dynkin diagram matches that of an $F_4$ non-simply laced group, as suggested when deleting the affine node.
\begin{table}[h!]
\begin{center}   
\begin{tabular}{|c|c|c|}\cline{2-3}
\multicolumn{1}{c|}{}& $Covering Theory$ & $Quotient$ \\ \hline
G: & $ (E_7 \times SU(2))\mathbb{Z}_2$& $F_4 \times \mathbb{Z}_2$ \\ \hline
$(h^{1,1},h^{2,1})$&$(19,19)$&$ (11,11)$ \\ 
$H_c$: & $ \mathbf{133}\oplus \mathbf{3}$   &$ \mathbf{52}$ \\
$T_{(1,0)}: $&$ 9$&$ 1$ \\  
$T_{(2,0)}:$ & $0$ &$ 4\times \mathcal{A}_1$  \\ \hline
\end{tabular}  
\end{center}
\caption{
\label{tab:FoldingQuotient} {\it Summary of the F-theory spectrum quotient group, with residual non-simply laced gauge $F_4$, appearing from a $\mathbb{Z}_2$ quotient of a Schoen threefold.}
}
\end{table} 
Upon taking the quotient, the curves $a_1=0$ and $b_1=0$ stay genus one curves with self intersection zero. Hence the former curve hosts an adjoint of $F_4$ and while the latter one does not contribute discrete charged singlets from the broken $SU(2)$ as argued in previous sections. Including the full superconformal matter sector, this spectrum is fully consistent with anomaly cancellation.
 
\subsection{Comments on bounds on discrete symmetries}
\label{sec:Comments}
In the previous Sections we have constructed numerous non-simply connected genus-one fibrations with n-sections, and due to the simple nature of quotienting CY threefolds, whose geometry and physics is fully specified by elliptic fibrations with finite Mordell-Weil group of order n. As a result of this relationship, the quotienting process provides a systematic way to construct 6-dimensional theories with $\mathbb{Z}_n$ discrete gauge symmetries.

This raises the natural question --- \emph{what discrete symmetries can appear in this context? Is there a bound on the order of the discrete groups?} In the context of the so-called ``Swampland program" \cite{Vafa:2005ui} it is of interest to map out what effective theories are realizable within F-theory in $6$-dimensions and in particular, to ask what is the maximal order of a discrete gauge symmetry?
 
From the constructions given here, it is clear that a classification of non-simply connected CY threefolds (and their multiple fibers) would have implications on the possible bounds for discrete symmetries and their link to superconformal matter. Also, from a related but complementary point of view a classification of Mordell-Weil torsion groups of CY threefolds would also be valuable for this question. However, at present neither type of classification yet exists within the literature. 

At present, the only classifcations of freely acting discrete symmetries of CY threefolds exist for specific datasets of manifold (and assume a coordinate action inherited from a simple ambient space). These include \cite{Braun:2010vc} for CICY threefolds and \cite{Braun:2017juz} for toric hypersurfaces. An analysis of the former has been undertaken to determine which symmetries are consistent with fibration structures \cite{Anderson:2018kwv} (based on the tools and classification in \cite{Anderson:2017aux,Anderson:2016ler,Anderson:2016cdu,Anderson:2015iia}). Although quotients by non-Abelian discrete groups are known for CY threefolds. In that work it was also found that in the set of CICY threefolds, only Abelian discrete groups preserve genus one fibration. Moreover the possible groups/orders appearing in that dataset are found to be
\begin{align}
\mathbb{Z}_n\,,   n\in\{2,3,4,6\}\, , \quad \mathbb{Z}_{2}\times \mathbb{Z}_m \, , m \in \{ 2,3,4\} \, , \quad \mathbb{Z}_3 \times \mathbb{Z}_3 \, .
\end{align}

Similarly to the discussion of MW torsion in the previous section, it is worth noting that $\mathbb{Z}_6$ is the highest order (single factor) appearing and $\mathbb{Z}_3 \times \mathbb{Z}_3$ the highest order product\footnote{Note however that clearly some discrete actions in the symmetry classifications listed above are missing since for example the Schoen threefold appears in the CICY threefold list \cite{Braun:2010vc}, but the $\mathbb{Z}_5$ symmetry (described in the previous section) is not inherited from a simple toric/projective ambient space.}.

It is also interesting to compare the discrete symmetries known above to those appearing in other constructions. Another relation between F-theory models with Mordell-Weil torsion and multi-sections also appeared in the context of applying \emph{fiber-wise} mirror symmetry of an elliptic fibration  \cite{Klevers:2014bqa,Oehlmann:2016wsb,Cvetic:2016ner} where it was observed that genus-one geometries and those with torsional sections were exchanged. This construction is only a statement about the structure of the generic fiber itself and no action on the base twofold was given\footnote{It has been shown in \cite{Huang:2018vup} that mirror symmetry on a full threefold can factorize into a fiber and base part, yielding also to the observed mirror structure of the generic fiber.}. Hence this construction gives further evidence that a systematic classification of torsion groups of elliptic fibrations could lead to a more systematic understanding of genus-one fibrations with multi-sections of the same order and hence discrete symmetries in F-theory. In \cite{Cvetic:2016ner} this connection was further related to mirror-symmetry in the context of Heterotic/F-theory duality \cite{Berglund:1998ej} of K3 surfaces that are stable degenerated. In that context discrete symmetries, torsional sections and subgroups of $E_8$ (and hence of bounded rank), naturally arise.

The observations above also appear to agree with recent classifications of $U(1)$ charges in the type IIB context \cite{Collinucci:2018aho,Collinucci:2019fnh,Cianci:2018vwv} using matrix factorization techniques, that are bounded to be not higher than charge six and hence upon Higgsing, there is a maximal $\mathbb{Z}_6$ symmetry over a generic base\footnote{A similar important role of $E_8$ has been played in swampland bounds of U(1) symmetries \cite{Lee:2019skh}.}.

To summarize, it seems that via known constructions in $6$-dimensions, a $\mathbb{Z}_6$ discrete symmetry seems to be the maximal order appearing thus far. It should be noted however that the 6-dimensional YM coupled to SUGRA theories constructed in \cite{Raghuram:2018hjn} seem to go beyond these bounds but do not have a full F-theory realization in their present form. Originating from Higgsed exotic representations of some non-Abelian groups, Abelian gauge group remnants, with up to $q=21$ U(1) charged singlets have been obtained that can potentially be broken further to a $\mathbb{Z}_{21}$ symmetry. Whether or not these fully broken discrete models exist and if they fit into the above picture is an interesting avenue for future research.
%%%%%%%%%%%%%%%%%%%%%

\section{Conclusions and future directions}
\label{sec:conclusion}
In this article we generalize the discussion of F-theory on smooth genus-one fibered Calabi-Yau threefold quotients initiated in \cite{Anderson:2018heq}. In particular, in contrast to that work, we focus here on freely acting discrete symmetries appearing on \emph{elliptically fibered} CY threefolds (frequently with multiple or torsional sections). In the F-theory physics of the $6$-dimensional theory, we find that the matter content of the downstairs quotient theory can be easily determined from the form of the upstairs covering space geometry and that moreover the induced symmetry actions on covering space divisors have clear ramifications for the number of tensor, hyper and vector multiplets in the downstairs theory. We have found examples of quotient CY threefolds with Abelian and non-Abelian non-simply connected gauge symmetries all of which must include (2,0) strongly coupled sectors gauged under the discrete symmetries. Geometrically this sector originates from points in the base, where the quotient acts like an orbifold and a free shift of the sections in the fiber of the F-theory torus producing a {\it multiple fiber}. Over these points, all sections are identified producing a genus-one geometry that results in a discrete gauge symmetry in $6$-dimensions. The residual massless degrees of freedom in the downstairs theory can be determined from the covering space in full generality. 

It should be noted that the quotient construction explored here always leads to fibrations over singular base manifolds. The presence of these singularities means that this class of theories serves as a toolbox to systematically construct 6-dimensional supergravity theories coupled to discrete charged $\mathcal{A}_n$ (2,0) superconformal matter from elliptic fibrations with non-trivial Mordell-Weil groups. 

Moreover, the construction we have outlined in this work admits several starting points for future research. These include several subtle cases that could potentially be considered in more detail, such as quotients of gauge groups of type $SU(N \times M) /\mathbb{Z}_N$  or $U(1)\times G/\mathbb{Z}_n$ that potentially lead to interesting residual gauge groups after quotienting. These groups arise from the existence of multisections rather than being unique to quotient constructions (indeed, these effects can appear over generic bases and a detailed analysis of those geometries is left for future research \cite{Anderson19}.)

In addition, even for the class of theories explored here it seems to be puzzling, from a field theory perspective, why the construction of $SU(n)/\mathbb{Z}_n$ groups should be constrained or forbidden at all. The fact that these may be bounded by the order of MW torsion would be interesting to understand from the point of view of coupling YM theories to SUGRA in $6$-dimensions. This either could point towards the realization of other more unconventional fibers with higher rank torsion groups or possibly be ruled out by more subtle anomalies. 

Finally, it would be interesting to see how the discrete symmetries studied in this work interact with more novel solutions of F-theory, including so-called ``T-brane" solutions \cite{Cecotti:2010bp,Anderson:2013rka,Anderson:2017rpr} and whether any bounds could be derived on the order of discrete symmetries or maximal charges of matter.
 
\section*{Acknowledgments}
P.K.O. would like to thank Fabio Apruzzi, Markus Dierigl, Mboyo Esole, Antonella Grassi, Sheldon Katz, Ling Lin and Fabian Ruehle  for valuable discussions. The work of L.A. and J.G. is supported in part by NSF
grant PHY-1720321.  The work of P.K.O. is supported by an individual DFG grant OE 657/1-1. The authors would like to gratefully acknowledge the hospitality of the Simons Center for Geometry and Physics (and the semester long program, \emph{The Geometry and Physics of Hitchin Systems}) during the completion of this work.
\newpage
\appendix

\section{Higher order torsion models}
\label{app:WSFtorsion}
This sections continues the more detailed discussion of F-theory of elliptic fibrations with higher order torsion and their quotients that has been started in Subsection~\ref{sec:Z5Quotient} with $\mathbb{Z}_5$. The explicit Weierstrass models have been constructed in  \cite{Aspinwall:1998xj} which we take here. For every model we show that it can be embedded into a Schoen manifold while avoiding non-flat fibers requires the base to be $dP_9$. Hodge numbers of the quotients are obtained from \cite{Bouchard:2007mf}.    
 
\subsection{The $\mathbf{\mathbb{Z}}_6$ torsion model}
The generic Weierstrass model with a $\mathbb{Z}_6$ torsion point is given as
\begin{align} 
\begin{split}
 y^2+a_1 xy &+\frac{1}{32}(a_1 - b_1)(3 a_1+ b_1)(a_1 + b_1) = x^3+\frac18(a_1 - b_1)(a_1+b_1) x^2\\
f=&\frac{1}{192} b_1 ( 3 a_1^3 - 3 a_1^2 b_1-3 a_1 b_1^2 - b_1^3 )   \\
g=&    \frac{1}{110592} (3 a_1^2 - 6 a_1 b_1-b_1^2)(9 a_1^4 -6  a_1^2 b_1^2 -24 a_1^3 b_1-11 b_1^4      ) \\
\Delta=& \frac{1}{2^{24}} (a_1-5 b_1)(3 a_1 + b_1)^2(a_1+b_1)^3(a_1 - b_1)^6  \, , \quad a_1 \in \mathcal{O}(K_b^{-1})\, , b_1 \in \mathcal{O}(K_b^{-1 })
\end{split}
\end{align}
Assuming that the sections $a_i$ and $b_i$ do not factorize further, this model admits an $SU(2) \times SU(3) \times SU(6)/\mathbb{Z}_6$ gauge group. Each gauge factor is localized over a genus one curve contributing a single adjoint hypermultiplet.
From the structure of the $\mathbb{Z}_6$ factor we do not expect bifundamental matter but at most trifundamentals. This is consistently reflected in the geometry where all three gauge group factors collide over the points $a_1 = b_1=0$ with multiplicity $n_{scp}=(K_b^{-1})^2$. From the gravitational anomaly, the number of complex structures and Kahler deformation are those of the Schoen manifold. Excluding all non-flat fibers over a $dP_9$ base allows to take a $\mathbb{Z}_6$ quotient. The details of the spectra are summarized in the following:
\begin{align}
\label{eq:Z6Summary}
\begin{array}{|c|c|c|}\cline{2-3}
\multicolumn{1}{c|}{}& $Cover Theory$ & $Quotient$ \\ \hline
G :& (SU(2)\times SU(3) \times SU(6))/\mathbb{Z}_6& \mathbb{Z}_6 \\ \hline
(h^{1,1},h^{2,1})&(19,19)& (3,3) \\ 
H_c :&  \mathbf{3}\oplus\mathbf{8}\oplus\mathbf{35}   & 0 \\
T_{(1,0)} & 9 & 1 \\  
T_{(2,0)}&0 & \mathcal{A}_1 \oplus \mathcal{A}_2 \oplus \mathcal{A}_5  \\ \hline
\end{array}
\end{align}

 \subsection{The $\mathbb{Z}_2 \times \mathbb{Z}_4$ model}
The $\mathbb{Z}_2 \times \mathbb{Z}_4$ WSF model is given as 
\begin{align}
\begin{split}
 y^2+a_1 xy &-(a_1)(b_1^2- \frac{1}{16}a_1^2 )y = x^3-(b_1^2 -\frac{1}{16} a_1) x^2\\
f=&-\frac{1}{798}  a_1^4-\frac{7}{24}a_1^2 b_1^2-\frac13 b_1^4   \\
g=&    \frac{1}{55296} (a_1^2 +16 b_1^2)(a_1^2-24 a_1 b_1 +16b_1^2)(a_1^2 +24 a_1 b_1 +16 b_1^2)
  \\
\Delta=& \frac{1}{2^{16}} a_1^2 b_1^2( a_1 - 4 b_1 )^4 (a_1 + 4 b_1)^4  \, , \quad a_1 \in \mathcal{O}(K_b^{-1})\, , b_1 \in \mathcal{O}(K_b^{-1 })
\end{split}
\end{align}
Again we find a generic $SU(2)^2 \times SU(4)^2$ gauge symmetry localized on genus 1 curves each. The torsion point forbids bifundamentals but requires in fact quad-fundamental representations which overshots the discriminant. Indeed over the $(K_b^{-1})^2$ points of collisions $a_1 = b_1=0$ we find an (4,6,12) points but leads to the expected Hodge numbers. Demanding the absence of these points requires again a $dP_9$ base which allows a quotient when the base is of the same torsion type. The spectra of covering and quotient theory are summarized as 
\begin{align}
\label{eq:Z2Z4Summary}
\begin{array}{|c|c|c|}\cline{2-3}
\multicolumn{1}{c|}{}& $Cover Theory$ & $Quotient$ \\ \hline
G :& (SU(2)\times SU(4))^2/\mathbb{Z}_2 \times \mathbb{Z}_4& \mathbb{Z}_2 \times \mathbb{Z}_4 \\ \hline
(h^{1,1},h^{2,1})&(19,19)& (3,3) \\ 
H_c :&  2\times (\mathbf{3}\oplus \mathbf{15})   & 0 \\
T_{(1,0)} & 9 & 1 \\  
T_{(2,0)}&0 &2\times( \mathcal{A}_1 \oplus \mathcal{A}_3 )  \\ \hline
\end{array}
\end{align}

\subsection{The $\mathbb{Z}_3 \times \mathbb{Z}_3$ models}
This Weierstrass model is given as 
\begin{align}
\begin{split}
 y^2+a_1 xy &-\frac13(a_1+w b_1)(a_1+ w^2 b_1) y b_1 = x^3-(a_1 - b_1)b_1 x^2 + \frac13(a_1 +w b_1)(a_1 + w^2 b_1) b_1^2 c \, ,\\
f=&-\frac{1}{48} a_1 (a_1- 2 b_1)(a_1 - 2w b_1)(a_1 -2w^2 b_1)    \\
g=&  \frac{1}{864}(a_1^2+2 a_1 b_1 - 2 b_1^2)(a_1^2+2w a_1 b_1 - 2w^2 b_1^2)(a_1^2+2 w^2 a_1 b_1-2w b_1^2)
  \\
\Delta=& \frac{1}{432}(a_1+b_1)^3(a_1 + w b_1)^3(a_1 + w^2 b_1)^3 b_1^3  \, , \quad a_1 \in \mathcal{O}(K_b^{-1})\, , b_1 \in \mathcal{O}(K_b^{-1 })\, , w = e^{\frac{2 \pi i }{3}}
\end{split}
\end{align}
Which indeed gives an $SU(3)^4$ gauge group. The two $\mathbb{Z}_3$ factors however forbid not only all bifundamental but also trifundamental representations and there is at most a quad-fundamental possible. This however overshots the discriminant and leads to (4,6,12) points which go away upon choosing a $dP_9$ base with the usual spectrum summarized as
\begin{align}
\label{eq:Z2Z4Summarya}
\begin{array}{|c|c|c|}\cline{2-3}
\multicolumn{1}{c|}{}& $Cover Theory$ & $Quotient$ \\ \hline
G :& (SU(2)\times SU(3))^4/\mathbb{Z}_3 \times \mathbb{Z}_3& \mathbb{Z}_3 \times \mathbb{Z}_3  \\ \hline
(h^{1,1},h^{2,1})&(19,19)& (3,3) \\ 
H_c :&  4\times  \mathbf{8}   & 0 \\
T_{(1,0)} & 9 & 1 \\  
T_{(2,0)}&0 &4\times \mathcal{A}_2   \\ \hline
\end{array}
\end{align}

% #################################
% #        Bibliography           #
% #################################
% \clearpage

\end{document}